\documentstyle [12pt]{article}
\begin{document}
\baselineskip 20.0 pt
\par
\mbox{}
\vskip -1.25in\par
\mbox{}
 \begin{flushright}
\makebox[1.5in][1]{UU-HEP-92/12}\\
\makebox[1.5in][1]{August 1992}
 \end{flushright}
 \vskip 0.25in

\begin{center}
{\bf On the KP Hierarchy, $\hat{W}_{\infty}$ Algebra,
and Conformal SL(2,R)/U(1) Model}\\
\vspace{5 pt}
{\bf II. ~ The Quantum Case}\\
\vspace{40 pt}
{Feng Yu and Yong-Shi Wu}\\
\vspace{20 pt}
{{\it Department of Physics, University of Utah}}\\
{{\it Salt Lake City, Utah 84112, U.S.A.}}\\
\vspace{40 pt}
{\large ABSTRACT}\\
\end{center}
\vspace{10 pt}

This paper is devoted to constructing a quantum version of
the famous KP hierarchy, by deforming its second Hamiltonian
structure, namely the nonlinear $\hat{W}_{\infty}$ algebra.
This is achieved by quantizing the conformal noncompact
$SL(2,R)_{k}/U(1)$ coset model, in which $\hat{W}_{\infty}$ appears
as a hidden current algebra. For the quantum $\hat{W}_{\infty}$ algebra
at level $k=1$, we have succeeded in constructing an infinite
set of commuting quantum charges in explicit
and closed form. Using them a completely integrable
quantum KP hierarchy is constructed in the Hamiltonian form.
A two boson realization of the quantum $\hat{W}_{\infty}$ currents
has played a crucial role in this exploration.

\newpage
\section{Introduction}
\setcounter{equation}{0}
\vspace{5 pt}

In the previous paper [1], we have extensively explored
the inter-relationship among the KP hierarchy [2],
nonlinear $\hat{W}_{\infty}$ algebra [3] and
conformal noncompact $SL(2,R)_{k}/U(1)$ coset model [4]
at the classical level. We have shown that the KP hierarchy and
$SL(2,R)_{k}/U(1)$ share a common $\hat{W}_{\infty}$
symmetry, which appears both as a (the second) Hamiltonian structure
in the former [5,3]
and a hidden current algebra in the latter [6]. Moreover, the
well-known set of infinitely many involutive KP conserved
charges give rise to an infinite $\hat{W}_{\infty}$ symmetry
in the $SL(2,R)_{k}/U(1)$ model, which keeps the $\hat{W}_{\infty}$
algebra invariant. Because of the connection of the coset model
to the black hole in 2D string theory [7], it is necessary
to study these issues at the quantum level. In particular, in this
paper we are interested in constructing a quantum version (or deformation)
of the KP hierarchy which remains completely integrable.

As is well-known, the essence of the complete integrability, either
at the classical or quantum level, is the existence of a complete
set of conserved charges (as many as basic variables) that are
mutually commuting and independent of each other (involutive).
This is known to be the case for the classical KP hierarchy [5]. The
main difficulty in obtaining a quantum integrable KP hierarchy is
to prove the existence of infinitely many involutive quantum charges.

To this end, it is natural to exploit the above-mentioned
inter-relationship among the KP hierarchy, $\hat{W}_{\infty}$
algebra and $SL(2,R)/U(1)$ model. Working in the Hamiltonian
formalism, quantization of the KP hierarchy is
reduced to deforming its classical Hamiltonian structure
and Hamiltonian functions. We choose to quantize the second KP
Hamiltonian structure (or the $\hat{W}_{\infty}$) which, as we have
shown in [1], appears naturally in the $SL(2,R)/U(1)$ model. Therefore
a consistent quantum deformation of $\hat{W}_{\infty}$ may be achieved
by quantizing the coset model. On the other hand, to deform the KP
Hamiltonian functions is a much harder problem, because there are no
general rules for doing this for highly nonlinear Hamiltonians.

To be more concrete, let us recall in brief some basic facts about
the classical KP hierarchy. It is an infinite set of evolution
equations in various times
$t_{m}$ $(m=1,2,\ldots)$ in the Lax form
\begin{eqnarray}
\frac{\partial L}{\partial t_{m}} = {[(L^{m})_{+}, L]}
\end{eqnarray}
where $L$ is the pseudo-differential operator
\begin{eqnarray}
L = D+\sum^{\infty}_{r=0}u_{r}D^{-r-1},~~~~~~~D\equiv \partial/\partial{z},
\end{eqnarray}
with coefficients $u_{r}$ being functions of $z$ and $t_{m}$;
and $(L^{m})_{+}$ denotes the differential part of $L^{m}$.
A fundamental property of the hierarchy (1.1) is that it is
Hamiltonian, namely it can be put into the form
\begin{eqnarray}
\frac{\partial u_{r}(z)}{\partial t_{m}} = {\{u_{r}(z), \oint_{0}H_{m+1}(w)
dw\}}.
\end{eqnarray}
Here the Hamiltonian functions are given by
\begin{eqnarray}
H_{m+1} = \frac{1}{m}Res~L^{m}
\end{eqnarray}
with $Res L^{m}$ standing for the coefficient of the $D^{-1}$
term in $L^{m}$. And the brackets are the Poisson ones
\begin{eqnarray}
{\{u_{r}(z), u_{s}(w)\}} = k_{rs}(z)\delta(z-w)
\end{eqnarray}
with the two choices of $k_{rs}$ explicitly given in ref.[3] or [1],
which respectively define the first [8] and second [5] KP Hamiltonian
structures. In this paper, we will focus only on the second KP
Hamiltonian structure, the nonlinear classical $\hat{W}_{\infty}$
algebra, from which the first Hamiltonian structure $W_{1+\infty}$
[8,9,10] can be obtained by appropriate contraction [5,1].

Eqs.(1.3)-(1.5) imply that the infinite number of independent
charges, $Q_{m} \equiv \oint H_{m}(z)dz$, are both conserved and
in involution:
\begin{eqnarray}
{\{ Q_{n}, Q_{m+1} \}} = \frac{\partial Q_{n}}{\partial t_{m}} = 0.
\end{eqnarray}

To obtain the quantum KP hierarchy, we need to seek for a quantum
deformation of the $\hat{W}_{\infty}$ algebra (1.5). Unlike the
case of (linear) Lie algebras, quantizing the nonlinear algebra
$\hat{W}_{\infty}$ is a quite nontrivial task: Simply changing the
Poisson brackets to commutators does not ensure closure of the
algebra. Fortunately, we have at hand an elegant free field
realization of the classical $\hat{W}_{\infty}$, which naturally
appears in the $SL(2,R)_{k}/U(1)$ model [6,1]. Hence quantization
of the coset model in the free field description [11] is
expected to yield a consistent quantum $\hat{W}_{\infty}$ algebra
in terms of the currents in the model. As we will see, the quantum
$\hat{W}_{\infty}$ obtained in this
manner receives intriguing deformations. The current-current
commutators acquire both nonvanishing central terms
and additional linear and nonlinear terms, and furthermore these
terms in general depend on the level $k$, an essential parameter
for the quantized coset model. This makes it very difficult to
construct the desired [12,6] set of infinitely many commuting quantum
$\hat{W}_{\infty}$ charges $Q_{m}$, which may be used to generate
the quantum KP hierarchy.  However, as reported in ref.[13],
such quantum $\hat{W}_{\infty}$-charges do exist, at least, at the
level $k=1$. In this paper we will give a rigorous proof for
their explicit construction.

This paper is organized as follows. In section 2, we generate the
quantum $\hat{W}_{\infty}$ currents, by means of the operator
product expansion (OPE), from the quantized $SL(2,R)_{k}$ $/U(1)$
currents, with the help of their free boson realization.
These currents are shown to form a closed algebra, which is a quantum
deformation of $\hat{W}_{\infty}$ and denoted as
$\hat{W}_{\infty}(k)$. Some very useful OPE's for the
quantum $\hat{W}_{\infty}$ currents are derived. Section 3 is
then devoted to explicit construction of an infinite number of
commuting quantum $\hat{W}_{\infty}$ charges in the case with
level $k=1$, by means of conformal field theory techniques.
In section 4, we use these charges to generate a set of infinitely
many compatible quantum flows, which are then justified to be a
quantum deformation of the KP hierarchy and have a natural free
field realization in terms of two bosons.\footnote[1]{We make
the following remark for caution. The KP flows we are going to
quantize are flows in the space of functions {$u_r(z); r=0,1,2,
\ldots$} of one variable $z$. Their quantization is principally
different from the quantization of the original KP equation
considered as a flow in the space of a function of two variables.}

\section{Quantum $\hat{W}_{\infty}$ and Its Two Boson Realization}
\setcounter{equation}{0}
\vspace{5 pt}

Usually at the quantum level a (linear) Lie algebra may acquire a
central extension, due to normal ordering of operators. However,
for a nonlinear algebra, the emergence
of central terms alone in quantum corrections would violate
closure, unless they are accompanied by additional
(mostly nonlinear) terms. There are no general rules for writing down
the additional terms. So quantization of a nonlinear algebra has to be
done case by case. A free field realization, if available, would be
very helpful.



Fortunately for the classical $\hat{W}_{\infty}$, we have previously
established [6,1] a free field realization in terms of two bosons,
which in the KP basis reads
\begin{eqnarray}
L = D+\sum^{\infty}_{r=0}u_{r}D^{-r-1} = D+\bar{j}\frac{1}{D-(\bar{j}+j)}j
\end{eqnarray}
with $\bar{j}(z)=\bar{\phi}'(z)$, $j(z)=\phi'(z)$, the free boson
currents. Moreover, this realization appears naturally in the classical
conformal $SL(2,R)/U(1)$ coset model through the following product
expansion:
\begin{eqnarray}
\psi_{+}(z)\psi_{-}(z') =
\sum^{\infty}_{r=0}u_{r}(z)\frac{(z-z')^{r}}{r!}
\end{eqnarray}
with the parafermionic currents $\psi_{+}$, $\psi_{-}$ given by
\begin{eqnarray}
\psi_{+}=\bar{j}e^{\bar{\phi}+\phi}, ~~~~
\psi_{-}=je^{-\bar{\phi}-\phi}.
\end{eqnarray}
in terms of the two free bosons $\bar{\phi}$ and $\phi$.
(Here classically, without loss of generality, level $k$ is set to 1.)
The main observation to our success in quantizing $\hat{W}_{\infty}$
is that a consistent deformation for the $\hat{W}_{\infty}$
generators that necessarily leads to a closed algebra at the
quantum level should naturally follow from quantizing the conformal
coset model. (This observation was inspired by a work of
Bakas and Kiritsis [12].)

Thus, let us start with the bosonized quantum $SL(2,R)_{k}$ currents [11]
\begin{eqnarray}
J_{\pm} &=& \sqrt{\frac{k}{2}}e^{\pm\sqrt{\frac{2}{k}}\phi_{3}}
(\phi_{1}' \mp i\sqrt{1-\frac{2}{k}}\phi_{2}')
e^{\pm\sqrt{\frac{2}{k}}\phi_{1}}, \nonumber\\
J_{3} &=& -\sqrt{\frac{k}{2}}\phi_{3}',
\end{eqnarray}
where $\phi_{i} (i=1,2,3)$ denote three free bosons, and $k$ is the level
parameter of the quantized model.  Also we have set the Planck constant
$\hbar =1$. Otherwise, $k$ should be replaced by $k/\hbar$. (Note that
each quantum current $\phi_i$ ($i$=1,2,3) has the dimension
$\hbar^{1/2}$.) Hence, the parameter $p\equiv k^{-1}$ plays the role of
the Planck constant as the essential parameter in quantum
corrections (or quantum deformation).
In eq.(2.4), we have included necessary quantum corrections
in the currents. The classical limit is recovered by
first rescaling $\phi_{i}\rightarrow\sqrt{k}\phi_{i}$, $J_i\rightarrow
kJ_i$ and then letting $k\rightarrow\infty$ [1].
Now by gauging away the $U(1)$ current $J_{3}$ or simply
setting $\phi_{3}=0$, from (2.7) we obtain the following
parafermionic currents of the quantized $SL(2,R)_{k}/U(1)$ model [14]
\begin{eqnarray}
\psi_{+}(z;p) &=&
\frac{1}{2}[(1+\sqrt{1-2p})\bar{j}+(1-\sqrt{1-2p})j]
e^{\sqrt{p}(\bar{\phi}+\phi)}, \nonumber\\
\psi_{-}(z;p) &=&
\frac{1}{2}[(1-\sqrt{1-2p})\bar{j}+(1+\sqrt{1-2p})j]
e^{-\sqrt{p}(\bar{\phi}+\phi)}.
\end{eqnarray}
Here $\psi_{\pm}(z;p)\equiv J_{\pm}\sqrt{p}$, and we have written
the two bosons $\phi_{1}(z)$ and $\phi_{2}(z)$,
of the same signature, satisfying
\begin{eqnarray}
\phi_{i}(z)\phi_{j}(z') \sim \delta_{ij}\, log(z-z') ~~~~~i,j=1,2 ,
\end{eqnarray}
as a pair of complex bosons $\bar{\phi}(z)=(1/\sqrt{2})(\phi_{1}-i\phi_{2})$,
$\phi (z)=(1/\sqrt{2})(\phi_{1}+i\phi_{2})$. Correspondingly, their currents
$\bar{j}(z)$, $j(z)$ satisfy the standard OPE's
\begin{eqnarray}
\bar{j}(z)j(z') \sim \frac{1}{(z-z')^{2}}, ~~~~~
\bar{j}(z)\bar{j}(z') \sim j(z)j(z') \sim 0.
\end{eqnarray}
Generalizing the classical equation (2.2), we use the whole OPE (up to all
orders), $\psi_{+}(z)\psi_{-}(z')$, of the $SL(2,R)_{k}/U(1)$ currents (2.5)
\begin{eqnarray}
\psi_{+}(z;p)\psi_{-}(z';p)~ = ~
\epsilon^{-2p}\{\epsilon^{-2}+\sum^{\infty}_{r=0}u_{r}(z;p)
\frac{\epsilon^{r}}{r!}\}
\end{eqnarray}
(with $\epsilon\equiv z-z'$) to generates the quantum
$\hat{W}_{\infty}(p)$ generators $u_{r}(z;p)$ in the KP basis. A closed
expression for all $u_{r}(z;p)$ can be derived as follows.

Proposition 1: The expansion coefficients $u_{r}(p,z)$ in the OPE (2.8)
are given by
\begin{eqnarray}
u_{r}(z;p) &=& \frac{1}{4}\sum^{r}_{r_{k+1}=0}\sum^{r}_{r_{k}=r_{k+1}}
\sum^{r}_{r_{k-1}=r_{k}}\cdots \sum^{r}_{r_{1}=r_{2}} \nonumber\\
& & \frac{(-1)^{r-r_{k}+r_{k+1}}\sqrt{p}^{r_{k}-r_{k+1}}r!}{r_{k+1}!
(r_{k}-r_{k+1})!(r_{k-1}-r_{k}+1)!\cdots (r_{1}-r_{2}+1)!(r-r_{1}+1)!}
\nonumber\\
& & ~~\times ((1+\sqrt{1-2p})\bar{j}+(1-\sqrt{1-2p})j) \nonumber\\
& & ~~\times ((1-\sqrt{1-2p})\bar{j}+(1+\sqrt{1-2p})j)^{(r_{k+1})} \nonumber\\
& & ~~\times (\bar{j}+j)^{(r_{k-1}-r_{k})}\cdots (\bar{j}+j)^{(r_{1}-r_{2})}
(\bar{j}+j)^{(r-r_{1})} \nonumber\\
& & + \frac{(-1)^{r+1}\sqrt{p}}{(r+2)(r+1)}(\bar{j}+j)^{(r+1)} \nonumber\\
& & + \frac{(-1)^{r}\sqrt{p}}{2(r+1)}
((1-\sqrt{1-2p})\bar{j}+(1+\sqrt{1-2p})j)^{(r+1)} \nonumber\\
& & + \sum^{r}_{r_{k+2}=0}\sum^{r}_{r_{k+1}=r_{k+2}}
\cdots \sum^{r}_{r_{1}=r_{2}} \nonumber\\
& & \frac{(-1)^{r-r_{k+2}}\sqrt{p}^{r_{k+2}+2}r!}{(r_{k+2}+2)!
(r_{k+1}-r_{k+2}+1)!\cdots (r_{1}-r_{2}+1)!(r-r_{1}+1)!}
\nonumber\\
& & ~~\times (\bar{j}+j)^{(r_{k+1}-r_{k+2})}\cdots (\bar{j}+j)^{(r_{1}-r_{2})}
(\bar{j}+j)^{(r-r_{1})} \nonumber\\
& & + \frac{1}{2}\sum^{r}_{r_{k+1}=0}\sum^{r}_{r_{k}=r_{k+1}}
\cdots \sum^{r}_{r_{1}=r_{2}} \nonumber\\
& & \frac{(-1)^{r-r_{k+1}+1}\sqrt{p}^{r_{k+1}+2}r!}{(r_{k+1}+1)!
(r_{k}-r_{k+1}+1)!\cdots (r_{1}-r_{2}+1)!(r-r_{1}+1)!}
\nonumber\\
& & ~~\times ((1+\sqrt{1-2p})\bar{j}+(1-\sqrt{1-2p})j) \nonumber\\
& & ~~\times (\bar{j}+j)^{(r_{k}-r_{k+1})}\cdots (\bar{j}+j)^{(r_{1}-r_{2})}
(\bar{j}+j)^{(r-r_{1})} \nonumber\\
& & + \frac{1}{2}\sum^{r}_{r_{k+2}=0}\sum^{r}_{r_{k+1}=r_{k+2}}
\cdots \sum^{r}_{r_{1}=r_{2}} \nonumber\\
& & \frac{(-1)^{r-r_{k+1}+r_{k+2}+1}\sqrt{p}^{r_{k+1}-r_{k+2}+2}r!}{r_{k+2}!
(r_{k+1}-r_{k+2}+1)!\cdots (r_{1}-r_{2}+1)!(r-r_{1}+1)!}
\nonumber\\
& & ~~\times ((1-\sqrt{1-2p})\bar{j}+(1+\sqrt{1-2p})j)^{(r_{k+2})} \nonumber\\
& & ~~\times (\bar{j}+j)^{(r_{k}-r_{k+1})}\cdots (\bar{j}+j)^{(r_{1}-r_{2})}
(\bar{j}+j)^{(r-r_{1})} ,
\end{eqnarray}
where $j^{(r)} \equiv (\partial_{z})^r ~j$, etc.

We observe that the quantum $\hat{W}_{\infty}(p)$ generators (2.9)
depend only on the currents $\bar{j}$ and $j$, as in the classical case.
The normal ordering on the current operators on the right side of (2.9)
is understood. But we suppress the notation for convenience.
In case there is a confusion, we will put a {\it dot} symbol between two
operators to denote their standard OPE to all orders.

Lemma 1.
\begin{eqnarray}
& & e^{\sqrt{p}(\bar{\phi}+\phi)}(z)\cdot e^{-\sqrt{p}(\bar{\phi}+\phi)}
(z-\epsilon) ~=~ \epsilon^{-2p}e^{\sqrt{p}(\bar{\phi}+\phi)}(z)
e^{-\sqrt{p}(\bar{\phi}+\phi)}(z-\epsilon); \\
& & \bar{j}(z)\cdot e^{-\sqrt{p}(\bar{\phi}+\phi)}(z-\epsilon) ~=~
(\bar{j}-\epsilon^{-1}\sqrt{p})(z)e^{-\sqrt{p}(\bar{\phi}+\phi)}(z-\epsilon);
\\
& & e^{\sqrt{p}(\bar{\phi}+\phi)}(z)\cdot \bar{j}(z-\epsilon) ~=~
e^{\sqrt{p}(\bar{\phi}+\phi)}(z)(\bar{j}-\epsilon^{-1}\sqrt{p})(z-\epsilon);
\end{eqnarray}
and eqs.(2.14)-(2.15) hold for the $\bar{j}\leftrightarrow j$ interchange.

Proof: We show, e.g., eq. (2.11) by a straightforward calculation:
\begin{eqnarray}
& & \bar{j}(z)\cdot e^{-\sqrt{p}(\bar{\phi}+\phi)}(z-\epsilon) ~=~
\bar{j}(z)\cdot \sum^{\infty}_{m=0}\frac{(-\sqrt{p})^{m}}{m!}
(\bar{\phi}+\phi)^{m}(z-\epsilon) \nonumber\\
&=& \bar{j}(z)\sum^{\infty}_{m=0}\frac{(-\sqrt{p})^{m}}{m!}
(\bar{\phi}+\phi)^{m}(z-\epsilon)
+ \sum^{\infty}_{m=1}\frac{(-\sqrt{p})^{m}}{(m-1)!\epsilon}
(\bar{\phi}+\phi)^{m-1}(z-\epsilon) \nonumber\\
&=& \bar{j}(z)e^{-\sqrt{p}(\bar{\phi}+\phi)}(z-\epsilon)
-\epsilon^{-1}\sqrt{p}e^{-\sqrt{p}(\bar{\phi}+\phi)}(z-\epsilon) \nonumber
\end{eqnarray}
where we have used eq.(2.6). (QED)

Now the proof of Proposition 1: From Lemma 1 we have, for example,
\begin{eqnarray}
& & \bar{j}e^{\sqrt{p}(\bar{\phi}+\phi)}(z)\cdot
je^{-\sqrt{p}(\bar{\phi}+\phi)}(z-\epsilon) \nonumber\\
&=& \epsilon^{-2p}[\bar{j}e^{\sqrt{p}(\bar{\phi}+\phi)}(z)
je^{-\sqrt{p}(\bar{\phi}+\phi)}(z-\epsilon) + \epsilon^{-2}(1+p)
e^{\sqrt{p}(\bar{\phi}+\phi)}(z)e^{-\sqrt{p}(\bar{\phi}+\phi)}(z-\epsilon)
\nonumber\\
&-& \epsilon^{-1}\sqrt{p}\bar{j}e^{\sqrt{p}(\bar{\phi}+\phi)}(z)
e^{-\sqrt{p}(\bar{\phi}+\phi)}(z-\epsilon) - \epsilon^{-1}\sqrt{p}
e^{\sqrt{p}(\bar{\phi}+\phi)}(z)je^{-\sqrt{p}(\bar{\phi}+\phi)}(z-\epsilon)].
\end{eqnarray}
The operator product of two parafermion currents
\begin{eqnarray}
I(p) &\equiv& \psi_{+}(z;p)\cdot\psi_{-}(z-\epsilon;p) \nonumber\\
&=& \frac{1}{4} ((1+\sqrt{1-2p})\bar{j}+(1-\sqrt{1-2p})j)
e^{\sqrt{p}(\bar{\phi}+\phi)}(z) \nonumber\\
& & \cdot ((1-\sqrt{1-2p})\bar{j}+(1+\sqrt{1-2p})j)
e^{-\sqrt{p}(\bar{\phi}+\phi)}(z-\epsilon)
\end{eqnarray}
then becomes, after some reorganizations,
\begin{eqnarray}
I(p) &=& \frac{1}{4}\epsilon^{-2p}[((1+\sqrt{1-2p})\bar{j}+(1-\sqrt{1-2p})j)
e^{\sqrt{p}(\bar{\phi}+\phi)}(z) \nonumber\\
& & ~~\times ((1-\sqrt{1-2p})\bar{j}+(1+\sqrt{1-2p})j)
e^{-\sqrt{p}(\bar{\phi}+\phi)}(z-\epsilon) \nonumber\\
& & -2\epsilon^{-1}\sqrt{p}((1+\sqrt{1-2p})\bar{j}+(1-\sqrt{1-2p})j)
e^{\sqrt{p}(\bar{\phi}+\phi)}(z)e^{-\sqrt{p}(\bar{\phi}+\phi)}(z-\epsilon)
\nonumber\\
& & -2\epsilon^{-1}\sqrt{p}e^{\sqrt{p}(\bar{\phi}+\phi)}(z)
((1-\sqrt{1-2p})\bar{j}+(1+\sqrt{1-2p})j)
e^{-\sqrt{p}(\bar{\phi}+\phi)}(z-\epsilon) \nonumber\\
& & +4\epsilon^{-2}e^{\sqrt{p}(\bar{\phi}+\phi)}(z)
e^{-\sqrt{p}(\bar{\phi}+\phi)}(z-\epsilon) ].
\end{eqnarray}
By expanding eq.(2.15) in powers of $\epsilon$, and separating
the terms with powers of $\epsilon^{-2p-2}$ and $\epsilon^{-2p-1}$ from
the rest, it follows that
\begin{eqnarray}
I(p) &=& \epsilon^{-2p}[\epsilon^{-2}+ \frac{1}{4}
((1+\sqrt{1-2p})\bar{j}+(1-\sqrt{1-2p})j) \nonumber\\
& & ~~\times \sum_{k=0}^{\infty}\frac{\sqrt{p}^{k}}{k!}
(\sum_{m=0}^{\infty}\frac{(-1)^{m}}{(m+1)!}(\bar{j}+j)^{(m)}
\epsilon^{m+1})^{k} \nonumber\\
& & ~~\times \sum_{n=0}^{\infty}\frac{(-1)^{n}}{n!}
((1-\sqrt{1-2p})\bar{j}^{(n)}+(1+\sqrt{1-2p})j^{(n)})\epsilon^{n}
\nonumber\\
& & -\sqrt{p}\sum_{m=0}^{\infty}\frac{(-1)^{m}}{(m+2)!}(\bar{j}+j)^{(m+1)}
\epsilon^{m} \nonumber\\
& & +\frac{1}{2}\sqrt{p}\sum_{n=0}^{\infty}\frac{(-1)^{n}}{(n+1)!}
((1-\sqrt{1-2p})\bar{j}^{(n+1)}+(1+\sqrt{1-2p})j^{(n+1)})\epsilon^{n}
\nonumber\\
& & + \sum_{k=0}^{\infty}\frac{\sqrt{p}^{k+2}}{(k+2)!}
(\sum_{m=0}^{\infty}\frac{(-1)^{m}}{(m+1)!}(\bar{j}+j)^{(m)}
\epsilon^{m})^{k+2} \epsilon^{k} \nonumber\\
& & -\frac{1}{2}((1+\sqrt{1-2p})\bar{j}+(1-\sqrt{1-2p})j) \nonumber\\
& & ~~\times \sum_{k=0}^{\infty}\frac{\sqrt{p}^{k+2}}{(k+1)!}
(\sum_{m=0}^{\infty}\frac{(-1)^{m}}{(m+1)!}(\bar{j}+j)^{(m)}
\epsilon^{m})^{k+1} \epsilon^{k} \nonumber\\
& & -\frac{1}{2}\sum_{n=0}^{\infty}\frac{(-1)^{n}}{n!}
((1-\sqrt{1-2p})\bar{j}^{(n)}+(1+\sqrt{1-2p})j^{(n)})\epsilon^{n}
\nonumber\\
& & ~~\times \sum_{k=0}^{\infty}\frac{\sqrt{p}^{k+2}}{(k+1)!}
(\sum_{m=0}^{\infty}\frac{(-1)^{m}}{(m+1)!}(\bar{j}+j)^{(m)}
\epsilon^{m})^{k+1} \epsilon^{k} ].
\end{eqnarray}
We observe that the coefficient of $\epsilon^{-2p-2}$ term is unity and
all $\epsilon^{-2p-1}$ terms cancel against each other;
thus we have verified the powers in the expression (2.8).
Now we change the ordering of multi-summations to collect terms
with the same power in $\epsilon$ together. Then one can easily read off
eq.(2.9) from the resulting expression of $I(p)$. (QED)

Let us introduce the quantum KP operator as
\begin{eqnarray}
L(z;p) = D + \sum_{r=0}^{\infty}u_{r}(z;p)D^{-r-1}
\end{eqnarray}
with $u_{r}(z;p)$ precisely given by eq.(2.9).
It was proved in our previous paper [1] that the bi-local classical
function $\bar{j}e^{\bar{\phi}+\phi}(z)
je^{-\bar{\phi}-\phi}(z-\epsilon)$ corresponds to the
pseudo-differential operator $\bar{j}\frac{1}{D-(\bar{j}+j)}j$.
Here we define a correspondence between
the quantum KP operator (2.17) and the OPE (2.8) parallel to the
classical case: The bilocal operator
$F(z,z')=\sum_{r}p_{r}(z)(z-z')^{r}/r!$
is said to correspond to a pseudo-differential operator
$P(z)=\sum_{r}p_{r}(z)D^{-r-1}$ of the same set of coefficient functions,
and denoted as $F(z,z') \Longleftrightarrow P(z)$.
The first two terms of eq.(2.16) can be reformulated as
\begin{eqnarray}
& & \epsilon^{-2}+ \frac{1}{4}((1+\sqrt{1-2p})\bar{j}+(1-\sqrt{1-2p})j)
e^{\sqrt{p}(\bar{\phi}+\phi)}(z) \nonumber\\
& & \times ((1-\sqrt{1-2p})\bar{j}+(1+\sqrt{1-2p})j)
e^{-\sqrt{p}(\bar{\phi}+\phi)}(z-\epsilon)  \Longleftrightarrow \nonumber\\
& & D+ \frac{1}{4}((1+\sqrt{1-2p})\bar{j}+(1-\sqrt{1-2p})j)
\frac{1}{D-\sqrt{p}(\bar{j}+j)} \nonumber\\
& & \times  ((1-\sqrt{1-2p})\bar{j}+(1+\sqrt{1-2p})j)
\end{eqnarray}
where we have set $\epsilon^{-2}\Longleftrightarrow D$. It is
obvious that eq.(2.18) is a simple deformation of eq.(2.1). The
remaining terms in eq.(2.16) represent additional quantum corrections.

The complete structure of the quantum $\hat{W}_{\infty}(p)$
algebra can be manifested by the OPE's between two currents
$u_{r}(z;p)$ and $u_{s}(w;p)$, as usual in conformal field theory.
In principle, these OPE's can be extracted from the following
OPE of four $SL(2,R)_{k}/U(1)$ parafermionic currents:
\begin{eqnarray}
L(z;p)L(w;p) \Longleftrightarrow
(\epsilon\sigma)^{2p}(\psi_{+}(z;p)\psi_{-}(z-\epsilon;p))
(\psi_{+}(w;p)\psi_{-}(w-\sigma;p)).
\end{eqnarray}
The closure of the quantum $\hat{W}_{\infty}(p)$ algebra is ensured
by the closure of the OPE's associated with the enveloping algebra
of the $SL(2,R)_{k}$ currents (2.4) in the neutral sector in
the conformal model. In fact, in the $SL(2,R)_{k}$ neutral sector,
currents are always combinations of the products of
$J_{+}J_{-}^{(n)}$, $J_{3}$ and their derivatives.
Imposing the $J_{3}=0$ constraint selects combinations of the products
of $\psi_{+}\psi_{-}^{(n)}$ and their derivatives only. That they are
closed within the $\hat{W}_{\infty}(p)$ currents $u_{r}(z;p)$
in accordance to eq.(2.8) is thus guaranteed by the closure of
the enveloping $SL(2,R)_{k}/U(1)$ algebra in the neutral sector.
Therefore we have achieved a consistent quantum
deformation of $\hat{W}_{\infty}$, with the help of
the free field realization of the $SL(2,R)_k/U(1)$ coset model.
However, despite the crucial role of the model in the construction,
once we write the algebra in the form of OPE's among $u_r$ and $u_s$,
the currents $u_r$ may be considered as independent of each other,
and the associativity of the OPE's automatically leads to closed
Jacobi identities, which do not require the use of the
two boson representation (2.9). What we have obtained is thus
justified to be a quantum version of the full $\hat{W}_{\infty}$
algebra with independent currents, which holds even beyond
the context of the conformal coset model.

For the purpose of illustration and later use, let us give some
explicit expressions for the quantum $\hat{W}_{\infty}(p)$. From eq.(2.9),
the first few quantum $\hat{W}_{\infty}(p)$ generators read
\begin{eqnarray}
u_{0}(p) &=& (1-2p)\bar{j}j-\frac{1}{2}\sqrt{1-2p}\sqrt{p}(\bar{j}'-j'),
\nonumber\\
u_{1}(p) &=& -\frac{1}{2}((1-p)\sqrt{1-2p}+1-2p)\bar{j}j' +\frac{1}{2}
((1-p)\sqrt{1-2p}-1+2p)\bar{j}'j \nonumber\\
& & -\frac{1}{2}p\sqrt{1-2p}(\bar{j}\bar{j}'-jj') +\frac{1}{12}(3\sqrt{1-2p}-1)
\sqrt{p}\bar{j}'' \nonumber\\
& & - \frac{1}{12}(3\sqrt{1-2p}+1)\sqrt{p}j''
+(1-\frac{3}{2}p)\sqrt{p}(\bar{j}j^{2}+\bar{j}^{2}j) \nonumber\\
& & +\frac{1}{6}p\sqrt{p}(\bar{j}^{3}+j^{3}), \nonumber\\
u_{2}(p) &=& \frac{1}{2}((1-p)\sqrt{1-2p}+1-2p)\bar{j}j'' -\frac{1}{2}
((1-p)\sqrt{1-2p}-1+2p)\bar{j}''j \nonumber\\
& & -\frac{1}{2}p\bar{j}'j' +\frac{1}{4}(2\sqrt{1-2p}-1)p\bar{j}'^{2}
\nonumber\\
& & -\frac{1}{4}(2\sqrt{1-2p}+1)pj'^{2}
+\frac{1}{2}p\sqrt{1-2p}(\bar{j}\bar{j}''-jj'')
\nonumber\\
& & -\frac{1}{12}(2\sqrt{1-2p}-1)\sqrt{p}\bar{j}''' +\frac{1}{12}
(2\sqrt{1-2p}+1)\sqrt{p}j''' \nonumber\\
& & -((1-p)\sqrt{1-2p}+2-3p)\sqrt{p}\bar{j}jj' \nonumber\\
& & +((1-p)\sqrt{1-2p}-2+3p)\sqrt{p}\bar{j}\bar{j}'j \nonumber\\
& & +((1-\frac{1}{2}p)\sqrt{1-2p}-1+\frac{3}{2}p)\sqrt{p}\bar{j}'j^{2}
\nonumber\\
& & -((1-\frac{1}{2}p)\sqrt{1-2p}+1-\frac{3}{2}p)\sqrt{p}\bar{j}^{2}j'
\nonumber\\
& & -\frac{1}{2}(\sqrt{1-2p}+1)p\sqrt{p}\bar{j}^{2}\bar{j}'
+\frac{1}{2}(\sqrt{1-2p}-1)p\sqrt{p}j^{2}j' \nonumber\\
& & +(1-p)p(\bar{j}j^{3}+\bar{j}^{3}j) +(2-\frac{5}{2}p)p\bar{j}^{2}j^{2}
+\frac{1}{4}p^{2}(\bar{j}^{4}+j^{4}).
\end{eqnarray}
We have checked that the quantum generators
$W_{i}(k)$ $(k=p^{-1})$ constructed in ref.[12] can be obtained under the
following basis transformation:
\begin{eqnarray}
W_{2}(k) &=& \frac{1}{(1-2p)}u_{0}(p), \nonumber\\
W_{3}(k) &=& -4u_{1}(p)-2u_{0}'(p), \nonumber\\
W_{4}(k) &=& 16u_{2}(p)+16u_{1}'(p)+\frac{16}{5}u_{0}''(p)
-\frac{16(6+5p)}{(16-17p)}(u_{0}u_{0})(p)
\end{eqnarray}
where the local product $(u_0u_0)$ is given by
\begin{eqnarray}
(u_{0}u_{0})(p,z) &=& \frac{1}{2}(1-2p)^{2}(\bar{j}j''+\bar{j}''j)
- \frac{p}{2}(1-2p)\bar{j}'j' \nonumber\\
& & +\frac{p}{4}(1-2p)(\bar{j}'^{2}+j'^{2})
-\frac{1}{6}(1-2p)\sqrt{1-2p}\sqrt{p}(\bar{j}'''-j''') \nonumber\\
& & - (1-2p)\sqrt{1-2p}\sqrt{p}\bar{j}j(\bar{j}'-j')
+(1-2p)^{2}\bar{j}^{2}j^{2}.
\end{eqnarray}
Using these expressions, it is straightforward to calculate the
first few OPE's for the quantum $\hat{W}_{\infty}(p)$, from the OPE
(2.7) for the free boson currents. They read
\begin{eqnarray}
u_{0}(z)u_{0}(w) &=& (1-2p)(
\frac{2u_{0}(w)}{(z-w)^{2}}+\frac{u_{0}'(w)}{(z-w)}
+\frac{(1+p)}{(z-w)^{4}}) + O(z-w), \nonumber\\
u_{0}(z)u_{1}(w) &=& (1-2p)(
\frac{3u_{1}(w)}{(z-w)^{2}}+\frac{u_{1}'(w)}{(z-w)}
-\frac{2u_{0}(w)}{(z-w)^{3}} -\frac{2(1+p)}{(z-w)^{5}}) + O(z-w), \nonumber\\
u_{0}(z)u_{2}(w) &=& (1-2p)(
\frac{4u_{2}(w)}{(z-w)^{2}}+\frac{u_{2}'(w)}{(z-w)}
-\frac{6u_{1}(w)}{(z-w)^{3}} +\frac{2(3+p)u_{0}(w)}{(z-w)^{4}} \nonumber\\
& & +\frac{6(1+p)}{(z-w)^{6}}) + O(z-w),  \nonumber\\
u_{1}(z)u_{1}(w) &=& (1-\frac{3p}{2})
(\frac{4u_{2}(w)}{(z-w)^{2}}+\frac{2u_{2}'(w)}{(z-w)})
+(1-p)(\frac{2u_{1}'(w)}{(z-w)^{2}} +\frac{u_{1}''(w)}{(z-w)}) \nonumber\\
& & +\frac{8(1-p)pu_{0}(w)}{(z-w)^{4}}
+\frac{4(1-p)pu_{0}'(w)}{(z-w)^{3}} +\frac{pu_{0}''(w)}{(z-w)^{2}}
+\frac{(1+2p)pu_{0}'''(w)}{6(z-w)} \nonumber\\
& & +\frac{2p(u_{0}u_{0})(w)}{(z-w)^{2}}
+\frac{p(u_{0}u_{0})'(w)}{(z-w)}
-\frac{4(1+p)(3-8p+2p^{2})}{3(z-z')^{6}} \nonumber\\
& & +O(z-w).
\end{eqnarray}

{}From eq.(2.23) we see that the quantum $\hat{W}_{\infty}(p)$ algebra is
really $p$-dependent. Its $p$-independent classical limit --
$\hat{W}_{\infty}$ -- is recovered by taking $p\rightarrow 0$
after rescaling $u_{r}\rightarrow u_{r}/p$ and ${[,]}\rightarrow
p{\{,\}}$. This is equivalent to the standard $\hbar\rightarrow 0$
limit (after setting $p\rightarrow p\hbar$).
In this sense the $p=1$ case, corresponding to the
$SL(2,R)_{k}/U(1)$ coset model with level $k=1$, may be
interpreted as the ``typical'' quantum case.
For this value of $p$, a lot of expressions simplify and it becomes
possible to extract the most relevant $(z-w)^{-1}$ terms in the
OPE's between $u_{0}(z)$ or $u_{1}(z)$ and arbitrary $u_{s}(z)$.
In next section we will see these OPE's provide sufficient
information for constructing an infinite set of commuting
quantum $\hat{W}_{\infty}$ charges. (Note that to obtain
charge commutators from the OPE's, one needs a
double integrations over $z$ and $w$, so the terms in OPE
with other powers of $(z-w)^{-1}$ have no contribution to the
commutators.)

Proposition 2: In the case of $p=1$, we have
\begin{eqnarray}
u_{0}(z)u_{s}(w) &=& \frac{-1}{z-w} u_{s}'(w) + {\it ~terms~in~other~
powers~of~}(z-w)^{-1}, \nonumber\\
u_{1}(z)u_{s}(w) &=& \frac{-2}{z-w} [\sum^{s}_{l=1}(-1)^{l}
\left( \begin{array}{c}
s\\l
\end{array} \right) (u_{0}^{(l)}u_{s-l}) +\frac{u_{s+1}'}{(s+1)}
+\frac{(-1)^{s}u_{0}^{(s+2)}}{(s+1)(s+2)}](w) \nonumber\\
& & +{\it ~terms~in~other~powers~of~}(z-w)^{-1}.
\end{eqnarray}

Proof: Instead of using the explicit expression (2.9) of $u_{s}(z;p)$, we
begin with the closed form (2.15). For the OPE $u_0(z)u_s(w)$
we can in fact obtain a general expression for arbitrary $p$.
Note that the total derivatives in $u_{0}(z;p)$ (see eq.(2.20))
will not contribute to the $(z-w)^{-1}$ terms in the OPE, so we have
\begin{eqnarray}
J_{1} &\equiv& u_{0}(z;p)\cdot [\epsilon^{-2}+\sum^{\infty}_{s=0}
u_{s}(w;p)\frac{\epsilon^{s}}{s!}] \nonumber\\
&=& (1-2p)\bar{j}j(z)\cdot [\frac{1}{4}
((1+\sqrt{1-2p})\bar{j}+(1-\sqrt{1-2p})j)
e^{\sqrt{p}(\bar{\phi}+\phi)}(w) \nonumber\\
& & \times ((1-\sqrt{1-2p})\bar{j}+(1+\sqrt{1-2p})j)
e^{-\sqrt{p}(\bar{\phi}+\phi)}(w-\epsilon) \nonumber\\
& & -\frac{1}{2}\epsilon^{-1}\sqrt{p}((1+\sqrt{1-2p})\bar{j}+(1-\sqrt{1-2p})j)
e^{\sqrt{p}(\bar{\phi}+\phi)}(w) \nonumber\\
& & \times e^{-\sqrt{p}(\bar{\phi}+\phi)}(w-\epsilon)
-\frac{1}{2}\epsilon^{-1}\sqrt{p}e^{\sqrt{p}(\bar{\phi}+\phi)}(w) \nonumber\\
& & \times ((1-\sqrt{1-2p})\bar{j}+(1+\sqrt{1-2p})j)
e^{-\sqrt{p}(\bar{\phi}+\phi)}(w-\epsilon) \nonumber\\
& & +\epsilon^{-2}e^{\sqrt{p}(\bar{\phi}+\phi)}(w)
e^{-\sqrt{p}(\bar{\phi}+\phi)}(w-\epsilon) ] \nonumber\\
& & +{\it ~terms~in~other~powers~of~}(z-w)^{-1}.
\end{eqnarray}
We extract from eq.(2.25) the $(z-w)^{-1}$
terms and then reorganize them into the fashion of eq.(2.15).
In this way, one obtains
\begin{eqnarray}
J_{1} &=& \frac{1}{z-w}(1-2p)\partial_{w}  [\frac{1}{4}
((1+\sqrt{1-2p})\bar{j}+(1-\sqrt{1-2p})j)
e^{\sqrt{p}(\bar{\phi}+\phi)}(w) \nonumber\\
& & \times ((1-\sqrt{1-2p})\bar{j}+(1+\sqrt{1-2p})j)
e^{-\sqrt{p}(\bar{\phi}+\phi)}(w-\epsilon) \nonumber\\
& & -\frac{1}{2}\epsilon^{-1}\sqrt{p}((1+\sqrt{1-2p})\bar{j}+(1-\sqrt{1-2p})j)
e^{\sqrt{p}(\bar{\phi}+\phi)}(w) \nonumber\\
& & \times e^{-\sqrt{p}(\bar{\phi}+\phi)}(w-\epsilon)
-\frac{1}{2}\epsilon^{-1}\sqrt{p}e^{\sqrt{p}(\bar{\phi}+\phi)}(w) \nonumber\\
& & \times ((1-\sqrt{1-2p})\bar{j}+(1+\sqrt{1-2p})j)
e^{-\sqrt{p}(\bar{\phi}+\phi)}(w-\epsilon) \nonumber\\
& & +\epsilon^{-2}e^{\sqrt{p}(\bar{\phi}+\phi)}(w)
e^{-\sqrt{p}(\bar{\phi}+\phi)}(w-\epsilon) ] \nonumber\\
& & +{\it ~terms~in~other~powers~of~}\frac{1}{z-w} \nonumber\\
&=& \frac{1}{z-w}\sum^{\infty}_{s=0}(1-2p)u_{s}'(w;p)\frac{\epsilon^{s}}{s!}
+{\it ~terms~in~other~powers~of~}(z-w)^{-1}.
\end{eqnarray}
Comparaing eq.(2.26) with (2.25), we obtain the first equation of (2.24).

The same scheme works for the proof of the second equation of (2.24).
According to eq.(2.20), when $p=1$, $u_{1}$ reads
\begin{eqnarray}
u_{1}(z) = -\frac{1}{2}\bar{j}j(\bar{j}+j)+\frac{1}{6}(\bar{j}^{3}+j^{3})
+ {\it ~total~derivatives}.
\end{eqnarray}
It is sufficient to address
\begin{eqnarray}
J_{2} &\equiv& u_{1}(z)\cdot [\epsilon^{-2}+\sum^{\infty}_{s=0}
u_{s}(w)\frac{\epsilon^{s}}{s!}] \nonumber\\
&=& (-\frac{1}{2}\bar{j}j(\bar{j}+j)+\frac{1}{6}(\bar{j}^{3}+j^{3}))(z)
\cdot \nonumber\\
& & [\frac{1}{4}((1+\sqrt{1-2p})\bar{j}+(1-\sqrt{1-2p})j)
e^{\sqrt{p}(\bar{\phi}+\phi)}(w) \nonumber\\
& & \times ((1-\sqrt{1-2p})\bar{j}+(1+\sqrt{1-2p})j)
e^{-\sqrt{p}(\bar{\phi}+\phi)}(w-\epsilon) \nonumber\\
& & -\frac{1}{2}\epsilon^{-1}\sqrt{p}((1+\sqrt{1-2p})\bar{j}+(1-\sqrt{1-2p})j)
e^{\sqrt{p}(\bar{\phi}+\phi)}(w) \nonumber\\
& & \times e^{-\sqrt{p}(\bar{\phi}+\phi)}(w-\epsilon)
-\frac{1}{2}\epsilon^{-1}\sqrt{p}e^{\sqrt{p}(\bar{\phi}+\phi)}(w) \nonumber\\
& & \times ((1-\sqrt{1-2p})\bar{j}+(1+\sqrt{1-2p})j)
e^{-\sqrt{p}(\bar{\phi}+\phi)}(w-\epsilon) \nonumber\\
& & +\epsilon^{-2}e^{\sqrt{p}(\bar{\phi}+\phi)}(w)
e^{-\sqrt{p}(\bar{\phi}+\phi)}(w-\epsilon) ] \nonumber\\
& & +{\it ~terms~in~other~powers~of~}(z-w)^{-1}.
\end{eqnarray}
It turns out that (we skip the tedious calculation)
\begin{eqnarray}
J_{2} &=& \frac{-2}{z-w}\sum^{\infty}_{s=0} [\sum^{s}_{l=1}(-1)^{l}
\left( \begin{array}{c}
s\\l
\end{array} \right) (u_{0}^{(l)}u_{s-l}) +\frac{u_{s+1}'}{(s+1)}
+\frac{(-1)^{s}u_{0}^{(s+2)}}{(s+1)(s+2)}](w)\frac{\epsilon^{s}}{s!}
\nonumber\\
& & +{\it ~terms~in~other~powers~of~}(z-w)^{-1},
\end{eqnarray}
which leads to the desired equation. (QED)

\vspace{30 pt}
\section{Involutive Quantum $\hat{W}_{\infty}$ Charges}
\setcounter{equation}{0}
\vspace{5 pt}

The quantum $\hat{W}_{\infty}$ algebra obtained in last section is a
quantum version (deformation) of the second Hamiltonian structure (1.5)
of the classical KP hierarchy (1.3).
We intend to use it as the Hamiltonian structure of
the yet-to-be established quantum KP hierarchy.
What more we need is an infinite set of commuting
quantum $\hat{W}_{\infty}$ charges,
whose densities are quantum deformation of the classical Hamiltonian
functions (1.4).

For generic $p$, it is not hard to construct the first few
quantum Hamiltonians [6,12]
\begin{eqnarray}
H_{2}(z;p) &=& \frac{1}{(1-2p)}u_{0}(z;p), \nonumber\\
H_{3}(z;p) &=& u_{1}(z;p)+\frac{1}{2}u_{0}'(z;p), \nonumber\\
H_{4}(z;p) &=& u_{2}(z;p)+u_{1}'(z;p)+\frac{(5+4p)}{15}u_{0}''(z;p)
+\frac{p}{(1-2p)}(u_{0}u_{0})(z;p).
\end{eqnarray}
Their charges $Q_m(p)\equiv \oint_0 H_m(z;p) dz$
($m=2,3,4$) indeed mutually commute.
In principle, such construction may successively continue to higher
orders with rapidly increasing labor and effort.
Fortunately, in the ``typical'' quantum case with $p=1$, we can exploit
the two general OPE (2.24) to explicitly construct an infinite number
of independent, commuting quantum $\hat{W}_{\infty}$ charges $Q_{m}$.

To start, let us first review in brief some useful facts for
the local product of several local operators in conformal field theory.
First the local product $(AB)(z)$ of two local operators $A(z)$
and $B(z)$ is the $(w-z)^0$ term in their OPE; namely
\begin{eqnarray}
(AB)(z) = \oint_{z} \frac{A(w)B(z)}{w-z} dw
\end{eqnarray}
in which the small contour of integration encircles $z$. The action of
the $z$-derivative still satisfies usual Leibniz rule:
 \begin{eqnarray}
\partial_{z} (AB)(z) = ((\partial_{z}A)B)(z) + (A\partial_{z}B)(z).
\end{eqnarray}
This local product is noncommutative, i.e. $(AB)(z)\neq (BA)(z)$;
however they differ from each other only by total derivatives:
\begin{eqnarray}
\oint_{0} (AB-BA)(z) dz =0.
\end{eqnarray}
According to the definition (3.2), the operator product of $C(z)$
with the local product $(AB)(w)$ is given by
\begin{eqnarray}
C(z)(AB)(w) = ((C(z)A(w))B(w)) + (A(w)(C(z)B(w))).
\end{eqnarray}
So, the multiple local product are generally nonassociative:
e.g., $(A(BC))-(A(BC)) \neq 0$; but one has the relation
\begin{eqnarray}
(A(BC))-(B(AC)) = ((AB)C)-((BA)C).
\end{eqnarray}
Finally we define the symmetric local product of $N$ local
operators to be the totally symmetrized sum of their multiple local products
taken from the left:
\begin{eqnarray}
\langle A_{1}A_{2}\cdots A_{N} \rangle = \frac{1}{N!}\sum_{P\{i\}}
(\cdots ((A_{i_{1}}A_{i_{2}})A_{i_{3}})\cdots A_{i_{N}})
\end{eqnarray}
where $P\{i\}$ denotes the summation over all possible permutations.

Now we proceed to construct an infinite set of involutive
quantum $\hat{W}_{\infty}$ charges (with $p=1$).
Let us assign a degree 1 to $\partial_z$ and $r+2$ to $u_r$. We
assume that the quantum charge-density $H_{m}(z)$
is homogeneous and of degree $m$, without loss of generality.
Furthermore we assume that the leading term of $H_m(z)$ which is
linear in the highest-spin current $u_{m-2}$
has nonvanishing coefficient, which can be normalized to unity.
This ensures the mutual independence of these charge densities.
Therefore the most general form for $H_m(z)$ is a linear combination
of multiple local products of the currents $u_r(z)$ and
their derivatives:
\begin{eqnarray}
H_{m}(z) = \sum_{l}\sum_{\{i,a\}}C_{i_{1}i_{2}\cdots i_{l}}^{a_{1}a_{2}
\cdots a_{l}}(m) (\cdots (u_{i_{1}}^{(a_{1})}u_{i_{2}}^{(a_{2})})\cdots
u_{i_{l}}^{(a_{l})}) (z), ~~~~m=2,3,\ldots
\end{eqnarray}
where $l$ ($=1,2,\cdots,[m/2]$) is the number of currents $u_{r}$
in the product (the maximal value of $l$ being the integral part of
$m/2$); $\{i,a\}$ stands for the set of all possible indices $i$'s
and $a$'s satisfying
$i_{1}+i_{2}+\cdots +i_{l}+a_{1}+a_{2}+\cdots +a_{l}$ $=m-2l$; and
$C_{i_{1}i_{2}\cdots i_{l}}^{a_{1}a_{2}\cdots a_{l}}(m)$ are constant
coefficients.
We want to determine these coefficients
so that the corresponding charges $Q_{m}\equiv \oint H_{m}(z)dz$
commute with each other:
\begin{eqnarray}
{[\oint_{0} H_{n}(z)dz , \oint_{0} H_{m}(w)dw]} = 0.
\end{eqnarray}

Proposition 3:  The commutativity with $Q_2$, i.e.,
\begin{eqnarray}
{[\oint_{0} H_{2}(z)dz , \oint_{0} H_{m}(w)dw]} = 0
\end{eqnarray}
is always satisfied.

Proof: Recall that for arbitrary local operators $A(z)$ and $B(z)$, we can
rewrite the commutator of their integrals in terms of their OPE:
\begin{eqnarray}
{[\oint_{0}A(z)dz, \oint_{0}B(w)dw]} = \oint_{0}dz\oint_{z}dw
{}~A(z)\cdot B(w),
\end{eqnarray}
through a continuous deformation of integration contour. Hence (note
$H_{2}$ has to be $u_0$ according to the homogeneity assumption)
\begin{eqnarray}
K \equiv {[\oint_{0} H_{2}(z)dz , \oint_{0} H_{m}(w)dw]} =
\oint_{0}dz\oint_{z}dw ~u_{0}(z)\cdot H_{m}(w). \nonumber
\end{eqnarray}
With arbitrary coefficients $C$'s in eq.(3.8) we have,
by applying eq.(3.5),
\begin{eqnarray}
K = \oint_{0}dz\oint_{z}dw \sum^{l}_{k=1}
\sum_{l}\sum_{\{i,a\}}C_{i_{1}i_{2}\cdots i_{l}}^{a_{1}a_{2}
\cdots a_{l}}(m) (\cdots (u_{i_{1}}^{(a_{1})}u_{i_{2}}^{(a_{2})})\cdots
(u_{0}(z)\cdot u_{i_{k}}^{(a_{k})}(w))\cdots u_{i_{l}}^{(a_{l})}) (w).
\nonumber
\end{eqnarray}
It follows that, by using the first equation of (2.24) and eq.(3.3),
\begin{eqnarray}
K &=& \oint_{0}dz\oint_{z}\frac{dw}{w-z} \sum^{l}_{k=1}
\sum_{l}\sum_{\{i,a\}}C_{i_{1}i_{2}\cdots i_{l}}^{a_{1}a_{2}
\cdots a_{l}}(m) (\cdots (u_{i_{1}}^{(a_{1})}u_{i_{2}}^{(a_{2})})\cdots
u_{i_{k}}^{(a_{k}+1)})\cdots u_{i_{l}}^{(a_{l})})(w) \nonumber\\
&=&  \oint_{0}dz\oint_{z}\frac{dw}{w-z} \partial_{w}
(\sum_{l}\sum_{\{i,a\}}C_{i_{1}i_{2}\cdots i_{l}}^{a_{1}a_{2}
\cdots a_{l}}(m) (\cdots (u_{i_{1}}^{(a_{1})}u_{i_{2}}^{(a_{2})})\cdots
u_{i_{l}}^{(a_{l})})(w)) \nonumber\\
&=& 0.
\end{eqnarray}
(QED)

Thus the first set of nontrivial equations in eq.(3.9) start with $n=3$:
\begin{eqnarray}
{[\oint_{0} H_{3}(z)dz , \oint_{0} H_{m}(w)dw]} = 0.
\end{eqnarray}
Here $H_{3}$ can only be $u_{1}$ plus a derivative of $u_{0}$;
the latter does not contribute to the charge $Q_3$. In the following
we will show that all the charge density $H_{m}(z)$ modulo total
derivatives, and thus all charges $Q_{m}$, are completely determined
by eq.(3.13) alone. Both amusingly and amazingly, as we will see later,
the so-determined charges $Q_{m}$ automatically commute with each other.

A direct step-by-step construction, with extensive use of
the second equation of (2.24) and eqs.(3.3)-(3.6), gives {\it uniquely}
the first seven charges from eq.(3.13) as follows
\begin{eqnarray}
& & Q_{2} = \oint u_{0}(z)dz, \nonumber\\
& & Q_{3} = \oint u_{1}(z)dz, \nonumber\\
& & Q_{4} = \oint (u_{2}-u_{0}u_{0})(z)dz, \nonumber\\
& & Q_{5} = \oint (u_{3}-6u_{0}u_{1})(z)dz, \nonumber\\
& & Q_{6} = \oint (u_{4}-12u_{0}u_{2}-12u_{1}u_{1}+8(u_{0}u_{0})u_{0})(z)dz,
\nonumber\\
& & Q_{7} = \oint (u_{5}-20u_{0}u_{3}-60u_{1}u_{2}+60(u_{0}u_{0})u_{1}
+60(u_{0}u_{1})u_{0})(z)dz, \nonumber\\
& & Q_{8} = \oint (u_{6}-30u_{0}u_{4}-120u_{1}u_{3}-90u_{2}u_{2}
+180(u_{0}u_{0})u_{2} +180(u_{0}u_{2})u_{0} \nonumber\\
& & ~~~~~~~~+360(u_{1}u_{0})u_{1}+360(u_{1}u_{1})u_{0}
-180((u_{0}u_{0})u_{0})u_{0})(z)dz.
\end{eqnarray}

Our key observation is that these charges $Q_{m}$ share the
following nice features:

(i) No term contains any derivative of $u_{r}$ at all, if all
local products of currents are chosen to start from the left.

(ii) For every $Q_{m}$, its coefficients conspire to result in
totally symmetric multiple (local) products of the currents involved.

(iii) All the seven charges are given by
\begin{eqnarray}
Q_{m} = \oint_{0} \sum_{l} \sum_{\{i\}}C_{i_{1}i_{2}\cdots i_{l}}(m)
\langle u_{i_{1}}u_{i_{2}}\cdots u_{i_{l}} \rangle (z)dz
\end{eqnarray}
with the elegant expression for the coefficients
\begin{eqnarray}
C_{i_{1}i_{2}\cdots i_{l}}(m) = \frac{(-1)^{l-1}(l-1)!(m-2)!}{d_{1}!d_{2}!
\cdots d_{k}!i_{1}!i_{2}!\cdots i_{l}!},
\end{eqnarray}
where for given number of currents in the product, $l$,
the summation is over all partitions $\{i_{k}\}$ of $m-2l$,
satisfying $i_{1}+i_{2}+\cdots +i_{l}=m-2l$; and here $d$'s denote
the degeneracies in the partition:
$0\leq i_{1}=i_{2}=\cdots =i_{d_{1}}$ $<i_{d_{1}+1}=\cdots
=i_{d_{1}+d_{2}}$ $<\cdots =i_{d_{1}+d_{2}+\cdots +d_{k}(=l)}$.
Note the coefficient of the leading linear term $u_{m-2}$
is unity as desired.

Now let us show that these are actually true for arbitrary $Q_m$:

Proposition 4: Eq.(3.16) provides a unique solution of the form
(3.15) to eq.(3.13) for arbitrary $m$.

Before proceeding to the proof, we need, for technical preparation,

Lemma 2:
\begin{eqnarray}
\oint_{0} \langle \underline{A}_{0}A_{1}\cdots A_{i}\cdots A_{N}
\rangle (z)dz =
\oint_{0} \langle A_{0}A_{1}\cdots A_{i}\cdots A_{N} \rangle (z)dz
\end{eqnarray}
where the {\it bar} under $A_{0}$ on the left side indicates
it does not join in the symmetrization:
\begin{eqnarray}
\langle \underline{A}_{0}A_{1}\cdots A_{i}\cdots A_{N} \rangle &\equiv&
\frac{1}{N!}\{ (\cdots ((A_{0}A_{1})A_{2})\cdots A_{N})
+ (\cdots ((A_{0}A_{2})A_{1})\cdots A_{N}) \nonumber\\
& & + {\it ~all~other~permutations~among~} 1,2,\ldots, N \}.
\end{eqnarray}
So the left side is symmetrized with respect to the $N$ indices
$(1,2,\cdots,N)$, while the right side is symmetrized with respect
to all $N+1$ indices by definition.

Proof: In fact, we have a stronger result:
\begin{eqnarray}
& & N(\cdots ((A_{0}A_{1})A_{2})\cdots A_{N}) +
A_{1}(\cdots ((A_{0}A_{2})A_{3})\cdots A_{N}) \nonumber\\
& & +(A_{1}A_{2})(\cdots ((A_{0}A_{3})A_{4})\cdots A_{N})
+ \cdots + (\cdots ((A_{1}A_{2})A_{3})\cdots A_{N-1})(A_{0}A_{N}) \nonumber\\
& & + {\it ~permutations~among~} 1,2,\ldots, N \nonumber\\
&=& \sum^{N}_{k=1}(\cdots ((A_{1}A_{2})A_{3})\cdots A_{0})A_{k})\cdots A_{N})
+ A_{0}(\cdots ((A_{1}A_{2})A_{3})\cdots A_{N}) \nonumber\\
& & + (A_{0}A_{N-1})(\cdots ((A_{1}A_{2})A_{3})\cdots A_{N-2})A_{N})
+ \cdots + (\cdots ((A_{0}A_{2})A_{3})\cdots A_{N-1})(A_{1}A_{N}) \nonumber\\
& & + {\it ~permutations~among~} 1,2,\ldots, N.
\end{eqnarray}
After integration, it yields eq.(3.17) by using eq.(3.4).

Consider, for example, the case with $N=3$. For the right
side of eq.(3.19), with repeated use of eq.(3.6), we have
\begin{eqnarray}
& & ((A_{0}A_{1})A_{2})A_{3} + ((A_{1}A_{0})A_{2})A_{3} +
((A_{1}A_{2})A_{0})A_{3} + A_{0}((A_{1}A_{2})A_{3}) \nonumber\\
& & +(A_{0}A_{2})(A_{1}A_{3}) + (1,2,3)~{\it perm.} \nonumber\\
&=& 2((A_{0}A_{1})A_{2})A_{3} + (A_{1}(A_{0}A_{2}))A_{3} -
(A_{0}(A_{1}A_{2}))A_{3} + ((A_{1}A_{2})A_{0})A_{3} \nonumber\\
& & +A_{0}((A_{1}A_{2})A_{3}) + (A_{0}A_{2})(A_{1}A_{3})
+ (1,2,3)~{\it perm.} \nonumber\\
&=& 2((A_{0}A_{1})A_{2})A_{3} + ((A_{0}A_{2})A_{1})A_{3} +
A_{1}((A_{0}A_{2})A_{3}) + (A_{1}A_{2})(A_{0}A_{3}) \nonumber\\
& & +(1,2,3)~{\it perm.}, \nonumber
\end{eqnarray}
which is just the left side of (3.19). The general validity
is proved by induction. (QED)

Now the proof of Proposition 4: From eq.(3.11), we see
(3.13) is equivalent to
\begin{eqnarray}
\oint_{0}dz\oint_{z}dw u_{1}(z)H_{m}(w) = 0. \nonumber
\end{eqnarray}
Applying the second equation of (2.24), and performing the $w$
integration, we rewrite it as
\begin{eqnarray}
& & \oint_{0} \sum_l \sum_{\{i\}}C_{i_{1}i_{2}\cdots i_{l}}(m)
\sum^{l}_{a=1}
\langle u_{i_{1}}\cdots u_{i_{a-1}} [\sum^{i_{a}}_{k=1}(-1)^{k}
\left( \begin{array}{c}
i_{a}\\k
\end{array} \right) (u_{0}^{(k)}u_{i_{a}-k}) \nonumber\\
& & +\frac{u_{i_{a}+1}'}{(i_{a}+1)}
+\frac{(-1)^{i_{a}}u_{0}^{(i_{a}+2)}}{(i_{a}+1)(i_{a}+2)}] u_{i_{a+1}}
\cdots u_{i_{l}} \rangle (z)dz = 0.
\end{eqnarray}
Note that each term has only one derivative
on one of the currents. To verify eq.(3.16), one
needs to show that all the terms in eq.(3.20) with the same
order of derivative on one of the $u_{r}$'s must cancel each other. In
particular, collecting the terms with a first-order derivative
we want to verify that
\begin{eqnarray}
& & \oint_{0} \sum_l \sum_{\{i\}}C_{i_{1}i_{2}\cdots i_{l}}(m)\sum^{l}_{a=1}
\langle u_{i_{1}}\cdots u_{i_{a-1}} [-i_{a}(u_{0}'u_{i_{a}-1}) \nonumber\\
& & +\frac{u_{i_{a}+1}'}{(i_{a}+1)}]
u_{i_{a+1}}\cdots u_{i_{l}} \rangle (z)dz = 0,
\end{eqnarray}
which can further be decomposed into,
\begin{eqnarray}
& & \oint_{0} \sum_{\{i\}_{l+1}}C_{i_{1}^{d_{1}}i_{2}^{d_{2}}
\cdots i_{k}^{d_{k}}}(m)\sum^{k}_{a=1} \frac{d_{a}}{i_{a}+1}
\langle u_{i_{1}}^{d_{1}}\cdots u_{i_{a}}^{d_{a}-1}
u_{i_{a}+1}' u_{i_{a+1}}^{d_{a+1}} \cdots u_{i_{k}}^{d_{k}}\rangle (z)dz =
\nonumber\\
& & \oint_{0} \sum_{\{j\}_{l}}C_{j_{1}^{c_{1}}j_{2}^{c_{2}}
\cdots j_{n}^{c_{n}}}(m)\sum^{n}_{a=1} c_{a}j_{a} \langle
u_{j_{1}}^{c_{1}}\cdots (u_{0}'u_{j_{a}-1})u_{j_{a}}^{c_{a}-1}
u_{j_{a+1}}^{c_{a+1}} \cdots u_{j_{n}}^{c_{n}} \rangle (z)dz
\end{eqnarray}
for each $l$. (For convenience, we will call $l$,
the number of currents in a product, the level of the term.
Do not confuse it with the level of the model.)
Here the $C$'s on the left side are level-$(l+1)$ coefficients and
those on the right side level-$l$ ones: $\sum d_{a} =l+1$,
$\sum d_{a} i_{a}=m-2(l+1)$, and $\sum c_{b} =l$, $\sum c_{a} j_{b}= m-2l$,
with $i_{1}<\cdots <i_{k}$, $j_{1}<\cdots <j_{n}$.

For terms on the left side having derivative on the
highest-spin current $u_{i_{k}}$, we integrate by parts and turn such
terms into those containing no derivative on $u_{i_{k}}$. Terms of
the latter property will be called irreducible; in the following
we will assume each term in a given partition of $m-2l$ has been turned
into irreducble in this way. To handle the local product
$(u_{0}'u_{j_{a}-1})$ in the middle of the right side of eq.(3.22),
which is symmetrized as a whole with other currents, one can use Lemma 2
to symmetrize all the currents invovled in a given partition.
Let us prove eq.(3.22) partition by partition.

We will do this by induction for both the level $l$ and the first
index $i_1$ in the partition.  First, on the left side of (3.22)
which is already symmetrized,
there is a term (after integration by parts) with the first index
$i_{1}=0$ and of the form
$u_{0}'u_{0}^{d_{1}-1}u_{i_{2}}^{d_{2}}\cdots u_{i_{k-1}}^{d_{k-1}}
u_{i_{k}}^{d_{k}-1}u_{i_{k}+1}$, with coefficient
\begin{eqnarray}
-\frac{d_{1}d_{k}}{i_{k}+1}C_{0^{d_{1}}i_{2}^{d_{2}}
\cdots i_{k}^{d_{k}}}(m).
\end{eqnarray}
Terms of the same partition on the right side appear as
\begin{eqnarray}
& & C_{0^{d_{1}-2}1i_{2}^{d_{2}}\cdots i_{k-1}^{d_{k-1}}i_{k}^{d_{k}-1}
(i_{k}+1)}(m)\langle (u_{0}'u_{0})u_{0}^{d_{1}-2}u_{i_{2}}^{d_{2}}\cdots
u_{i_{k-1}}^{d_{k-1}}u_{i_{k}}^{d_{k}-1}u_{i_{k}+1} \rangle \nonumber\\
& &  +(i_{2}+1) C_{0^{d_{1}-1}i_{2}^{d_{2}-1}(i_{2}+1)
i_{3}^{d_{3}}
\cdots i_{k-1}^{d_{k-1}}i_{k}^{d_{k}-1}(i_{k}+1)}(m) \nonumber\\
& & \times \langle (u_{0}'u_{i_{2}})u_{0}^{d_{1}-1}
u_{i_{2}}^{d_{2}-1}u_{i_{3}}^{d_{3}}
\cdots u_{i_{k-1}}^{d_{k-1}}u_{i_{k}}^{d_{k}-1}u_{i_{k}+1} \rangle
\nonumber\\
& & +\cdots +(i_{k-1}+1) C_{0^{d_{1}-1}i_{2}^{d_{2}}
\cdots i_{k-1}^{d_{k-1}-1}(i_{k-1}+1)i_{k}^{d_{k}-1}(i_{k}+1)}(m) \nonumber\\
& & \times \langle (u_{0}'u_{i_{k-1}})u_{0}^{d_{1}-1}u_{i_{2}}^{d_{2}}
\cdots u_{i_{k-1}}^{d_{k-1}-1}u_{i_{k}}^{d_{k}-1}u_{i_{k}+1} \rangle
\nonumber\\
& & +2(i_{k}+1) C_{0^{d_{1}-1}i_{2}^{d_{2}}
\cdots i_{k-1}^{d_{k-1}}i_{k}^{d_{k}-2}(i_{k}+1)^{2}}(m)
\langle (u_{0}'u_{i_{k}})u_{0}^{d_{1}-1}u_{i_{2}}^{d_{2}}
\cdots u_{i_{k-1}}^{d_{k-1}}u_{i_{k}}^{d_{k}-2}u_{i_{k}+1} \rangle
\nonumber\\
& & +(i_{k}+2) C_{0^{d_{1}-1}i_{2}^{d_{2}}
\cdots i_{k-1}^{d_{k-1}}i_{k}^{d_{k}-1}(i_{k}+2)}(m)
\langle (u_{0}'u_{i_{k}+1})u_{0}^{d_{1}-1}u_{i_{2}}^{d_{2}}
\cdots u_{i_{k-1}}^{d_{k-1}}u_{i_{k}}^{d_{k}-1} \rangle
\nonumber\\
&=& \frac{(-1)^{l-1}(l-1)!(m-2)!}{(d_{1}-1)!d_{2}!\cdots d_{k-1}!(d_{k}-1)!
(i_{2}!)^{d_{2}}\cdots (i_{k-1}!)^{d_{k-1}}(i_{k}!)^{d_{k}-1}(i_{k}+1)!}
\nonumber\\
& & \times [(d_{1}-1)\langle \underline{(u_{0}'u_{0})}u_{0}^{d_{1}-2}
u_{i_{2}}^{d_{2}} \cdots u_{i_{k-1}}^{d_{k-1}}u_{i_{k}}^{d_{k}-1}u_{i_{k}+1}
\rangle \nonumber\\
& & + d_{2}\langle \underline{(u_{0}'u_{i_{2}})}u_{0}^{d_{1}-1}
u_{i_{2}}^{d_{2}-1}u_{i_{3}}^{d_{3}}
\cdots u_{i_{k-1}}^{d_{k-1}}u_{i_{k}}^{d_{k}-1}u_{i_{k}+1} \rangle
\nonumber\\
& & +\cdots + d_{k-1}\langle \underline{(u_{0}'u_{i_{k-1}})}u_{0}^{d_{1}-1}
u_{i_{2}}^{d_{2}} \cdots u_{i_{k-1}}^{d_{k-1}-1}u_{i_{k}}^{d_{k}-1}
u_{i_{k}+1} \rangle \nonumber\\
& & +(d_{k}-1)\langle
\underline{(u_{0}'u_{i_{k}})}u_{0}^{d_{1}-1}u_{i_{2}}^{d_{2}}
\cdots u_{i_{k-1}}^{d_{k-1}}u_{i_{k}}^{d_{k}-2}u_{i_{k}+1} \rangle \nonumber\\
& & + \langle \underline{(u_{0}'u_{i_{k}+1})}u_{0}^{d_{1}-1}u_{i_{2}}^{d_{2}}
\cdots u_{i_{k-1}}^{d_{k-1}}u_{i_{k}}^{d_{k}-1} \rangle ] \nonumber\\
&=& \frac{(-1)^{l-1}l!(m-2)!}{(d_{1}-1)!d_{2}!\cdots d_{k-1}!(d_{k}-1)!
(i_{2}!)^{d_{2}}\cdots (i_{k-1}!)^{d_{k-1}}(i_{k}!)^{d_{k}-1}(i_{k}+1)!}
\nonumber\\
& & \times \langle \underline{u_{0}'}u_{0}^{d_{1}-1}u_{i_{2}}^{d_{2}}
\cdots u_{i_{k-1}}^{d_{k-1}}u_{i_{k}}^{d_{k}-1}u_{i_{k}+1} \rangle.
\end{eqnarray}
One can remove the {\it bar} under $u_{0}'$ in accordance to Lemma 2
so that eq.(3.24) is totally symmetrized.
Here we have assumed the validity of eq.(3.16) for
level $l$. The equality between (3.23) and the coefficient of (3.24) requires
exactly the validity of (3.16) with $i_1=0$ at level $l+1$.
We note that in the derivation of (3.24), there are some subtleties
about degeneracies: eq.(3.24) is written in the case that
$i_2$ in the first term on the left side of (3.24) is not equal to 1;
when it is one, we need to adjust the expressions.
We have checked that in every case we always get the right side of (3.24).

Furthermore, we need to verify that (after the above-mentioned integration by
parts) all the terms having derivative on $u_{i}$ with $i\neq 0$ on the left
side of eq.(3.22) cancel each other. Consider the terms of the form
$u_{i_{1}}'u_{i_{1}}^{d_{1}-1}u_{i_{2}}^{d_{2}}\cdots
u_{i_{k-1}}^{d_{k-1}}u_{i_{k}}^{d_{k}-1}u_{i_{k}+1}$, with coefficients
\begin{eqnarray}
-\frac{d_{1}d_{k}}{i_{k}+1}C_{i_{1}^{d_{1}}i_{2}^{d_{2}}\cdots
i_{k}^{d_{k}}}(m) +\frac{1}{i_{1}}C_{(i_{1}-1)i_{1}^{d_{1}-1}i_{2}^{d_{2}}
\cdots i_{k-1}^{d_{k-1}}i_{k}^{d_{k}-1}(i_{k}+1)}(m).
\end{eqnarray}
Assuming eq.(3.16) is true for the level-$(l+1)$ coefficient with the first
index $i_{1}-1$, the vanishing of (3.25) yields the correct
level-$(l+1)$ coefficient $C_{i_{1}^{d_{1}}i_{2}^{d_{2}}\cdots
i_{k}^{d_{k}}}(m)$ with the first index $i_{1}(\neq 0)$.

Similarly, the cancellation occurs for the terms with the derivative on
higher-spin currents $u_{i_{a}}$ (for $i_{a}\neq i_{1}, i_{k}$)
on the left hand side of (3.22), which are of the form
$u_{i_{1}}^{d_{1}}u_{i_{2}}^{d_{2}}\cdots u_{i_{a}}^{d_{a}-1}u_{i_{a}+1}'
u_{i_{a+1}}^{d_{a+1}}\cdots u_{i_{k-1}}^{d_{k-1}}u_{i_{k}}^{d_{k}-1}
u_{i_{k}+1}$:
\begin{eqnarray}
& & -\frac{d_{k}}{i_{k}+1}C_{i_{1}^{d_{1}}i_{2}^{d_{2}}\cdots
i_{a}^{d_{a}-1}(i_{a}+1)i_{a+1}^{d_{a+1}}\cdots i_{k}^{d_{k}}}(m) +
\nonumber\\
& & \frac{d_{a}}{i_{a}+1}C_{i_{1}^{d_{1}}i_{2}^{d_{2}}\cdots
i_{a}^{d_{a}}i_{a+1}^{d_{a+1}}\cdots
i_{k-1}^{d_{k-1}}i_{k}^{d_{k}-1}(i_{k}+1)}(m) = 0.
\end{eqnarray}
In the same manner, we have checked the validity of (3.22) for
all other partitions, of the general form
$u_{i_{1}}^{d_{1}}u_{i_{2}}^{d_{2}}\cdots
u_{i_{a}}^{d_{a}-1}u_{i_{a}+1}'u_{i_{a+1}}^{d_{a+1}}\cdots
u_{i_{k}}^{d_{k}}$.

With eq.(3.21) or (3.22) established, eq.(3.20) reduces to
\begin{eqnarray}
& & \oint_{0}\sum_l \sum_{\{i\}}C_{i_{1}i_{2}\cdots i_{l}}(m)\sum^{l}_{a=1}
\langle u_{i_{1}}\cdots u_{i_{a-1}} [\sum_{k=2}^{i_{a}} (-1)^{k}
\left( \begin{array}{c}
i_{a}\\k
\end{array} \right) (u_{0}^{(k)}u_{i_{a}-k}) \nonumber\\
& & +\frac{(-1)^{i_{a}}u_{0}^{(i_{a}+2)}}{(i_{a}+1)(i_{a}+2)}] u_{i_{a+1}}
\cdots u_{i_{l}} \rangle (z)dz = 0,
\end{eqnarray}
which is equaivalent to, at each level $l$,
\begin{eqnarray}
& & \sum_{\{j\}_{l}} C_{j_{1}^{c_{1}}j_{2}^{c_{2}}\cdots j_{n}^{c_{n}}}(m)
\sum^{n}_{a=1} c_{a} \langle u_{j_{1}}^{c_{1}}\cdots u_{j_{a-1}}^{c_{a-1}}
[\sum_{b=2}^{j_{a}} (-1)^{b}
\left( \begin{array}{c}
j_{a}\\b
\end{array} \right) (u_{0}^{(b)}u_{j_{a}-b})]
u_{j_{a}}^{c_{a}-1}u_{j_{a+1}}^{c_{a+1}}\cdots u_{j_{n}}^{c_{n}} \rangle =
\nonumber\\
& & \sum_{\{i\}_{l+1}} C_{i_{1}^{d_{1}}i_{2}^{d_{2}}\cdots i_{k}^{d_{k}}}(m)
\sum^{k}_{a=1} \frac{(-1)^{i_{a}+1}d_{a}}{(i_{a}+1)(i_{a}+2)}
\langle u_{0}^{(i_{a}+2)}u_{i_{1}}^{d_{1}}\cdots u_{i_{a-1}}^{d_{a-1}}
u_{i_{a}}^{d_{a}-1}u_{i_{a+1}}^{d_{a+1}}\cdots u_{i_{k}}^{d_{k}} \rangle,
\end{eqnarray}
where the indices are of the similar meaning as explained below eq.(3.22).
Again we need to separate terms having the same order of derivative
on one of the currents and show the cancellation among them.
Notice that all terms in eq.(3.28) are already irreducible. Consider
a generic partition:
$u_{0}^{(i_{a}+2)}u_{i_{1}}^{d_{1}}\cdots u_{i_{a-1}}^{d_{a-1}}
u_{i_{a}}^{d_{a}-1}u_{i_{a+1}}^{d_{a+1}}\cdots u_{i_{k}}^{d_{k}}$.
Only one term of such partition appears on the symmetrized
left side of (3.28),
which has the coefficient
\begin{eqnarray}
\frac{(-1)^{i_{a}+1}d_{a}}{(i_{a}+1)(i_{a}+2)}
C_{i_{1}^{d_{1}}i_{2}^{d_{2}}\cdots i_{k}^{d_{k}}}(m).
\end{eqnarray}
But from the right side, there emerge in total $k$ terms of the same
partition:
\begin{eqnarray}
& & (-1)^{i_{a}} \left( \begin{array}{c}
i_{a}+i_{1}+2 \\ i_{a}+2
\end{array} \right) C_{i_{1}^{d_{1}-1}(i_{1}+i_{a}+2)i_{2}^{d_{2}}\cdots
i_{a-1}^{d_{a-1}}i_{a}^{d_{a}-1}i_{a+1}^{d_{a+1}}\cdots i_{k}^{d_{k}}}(m)
\nonumber\\
& & \times \langle (u_{0}^{(i_{a}+2)}u_{i_{1}})u_{i_{1}}^{d_{1}-1}
u_{i_{2}}^{d_{2}} \cdots u_{i_{a-1}}^{d_{a-1}}u_{i_{a}}^{d_{a}-1}
u_{i_{a+1}}^{d_{a+1}}\cdots u_{i_{k}}^{d_{k}} \rangle \nonumber\\
& & + (-1)^{i_{a}} \left( \begin{array}{c}
i_{a}+i_{2}+2 \\ i_{a}+2
\end{array} \right) C_{i_{1}^{d_{1}}i_{2}^{d_{2}-1}(i_{2}+i_{a}+2)
i_{3}^{d_{3}}\cdots
i_{a-1}^{d_{a-1}}i_{a}^{d_{a}-1}i_{a+1}^{d_{a+1}}\cdots i_{k}^{d_{k}}}(m)
\nonumber\\
& & \times \langle
(u_{0}^{(i_{a}+2)}u_{i_{2}})u_{i_{1}}^{d_{1}}u_{i_{2}}^{d_{2}-1}
u_{i_{3}}^{d_{3}} \cdots u_{i_{a-1}}^{d_{a-1}}u_{i_{a}}^{d_{a}-1}
u_{i_{a+1}}^{d_{a+1}}\cdots u_{i_{k}}^{d_{k}} \rangle \nonumber\\
& & +\cdots + (-1)^{i_{a}} \left( \begin{array}{c}
2i_{a}+2 \\ i_{a}+2
\end{array} \right) C_{i_{1}^{d_{1}}\cdots i_{a-1}^{d_{a-1}}i_{a}^{d_{a}-2}
(2i_{a}+2)i_{a+1}^{d_{a+1}}\cdots i_{k}^{d_{k}}}(m) \nonumber\\
& & \times \langle (u_{0}^{(i_{a}+2)}u_{i_{a}})u_{i_{1}}^{d_{1}}\cdots
u_{i_{a-1}}^{d_{a-1}}u_{i_{a}}^{d_{a}-2}
u_{i_{a+1}}^{d_{a+1}}\cdots u_{i_{k}}^{d_{k}} \rangle \nonumber\\
& & +\cdots + (-1)^{i_{a}} \left( \begin{array}{c}
i_{a}+i_{k}+2 \\ i_{a}+2
\end{array} \right) C_{i_{1}^{d_{1}}\cdots i_{a-1}^{d_{a-1}}i_{a}^{d_{a}-1}
i_{a+1}^{d_{a+1}}\cdots i_{k-1}^{d_{k-1}}i_{k}^{d_{k}-1}(i_{k}+i_{a}+2)}(m)
\nonumber\\
& & \times \langle (u_{0}^{(i_{a}+2)}u_{i_{k}})u_{i_{1}}^{d_{1}}\cdots
u_{i_{a-1}}^{d_{a-1}}u_{i_{a}}^{d_{a}-1}u_{i_{a+1}}^{d_{a+1}}\cdots
u_{i_{k-1}}^{d_{k-1}}u_{i_{k}}^{d_{k}-1} \rangle \nonumber\\
&=& \frac{(-1)^{i_{a}+l-1}(l-1)!(m-2)!}{d_{1}!\cdots d_{a-1}!(d_{a}-1)!
d_{a+1}!\cdots d_{k}!
(i_{1}!)^{d_{1}}\cdots (i_{a-1}!)^{d_{a-1}}(i_{a}!)^{d_{a}-1}(i_{a}+2)!
(i_{a+1}!)^{d_{a+1}}\cdots (i_{k}!)^{d_{k}}}
\nonumber\\
& & \times [ d_{1}\langle
\underline{(u_{0}^{(i_{a}+2)}u_{i_{1}})}u_{i_{1}}^{d_{1}-1}
u_{i_{2}}^{d_{2}} \cdots u_{i_{a-1}}^{d_{a-1}}u_{i_{a}}^{d_{a}-1}
u_{i_{a+1}}^{d_{a+1}}\cdots u_{i_{k}}^{d_{k}} \rangle \nonumber\\
& & + d_{2}\langle \underline{(u_{0}^{(i_{a}+2)}u_{i_{2}})}u_{i_{1}}^{d_{1}}
u_{i_{2}}^{d_{2}-1}
u_{i_{3}}^{d_{3}} \cdots u_{i_{a-1}}^{d_{a-1}}u_{i_{a}}^{d_{a}-1}
u_{i_{a+1}}^{d_{a+1}}\cdots u_{i_{k}}^{d_{k}} \rangle \nonumber\\
& & +\cdots + (d_{a}-1)\langle \underline{(u_{0}^{(i_{a}+2)}u_{i_{a}})}
u_{i_{1}}^{d_{1}}\cdots u_{i_{a-1}}^{d_{a-1}}u_{i_{a}}^{d_{a}-2}
u_{i_{a+1}}^{d_{a+1}}\cdots u_{i_{k}}^{d_{k}} \rangle \nonumber\\
& & +\cdots + d_{k}\langle \underline{(u_{0}^{(i_{a}+2)}u_{i_{k}})}
u_{i_{1}}^{d_{1}}\cdots
u_{i_{a-1}}^{d_{a-1}}u_{i_{a}}^{d_{a}-1}u_{i_{a+1}}^{d_{a+1}}\cdots
u_{i_{k-1}}^{d_{k-1}}u_{i_{k}}^{d_{k}-1} \rangle ] \nonumber\\
&=& \frac{(-1)^{i_{a}+l-1}l!(m-2)!d_{a}}{d_{1}!\cdots d_{k}!
(i_{1}!)^{d_{1}}\cdots (i_{k}!)^{d_{k}}(i_{a}+1)(i_{a}+2)}
\langle \underline{u_{0}^{(i_{a}+2)}}u_{i_{1}}^{d_{1}}\cdots
u_{i_{a-1}}^{d_{a-1}}u_{i_{a}}^{d_{a}-1}u_{i_{a+1}}^{d_{a+1}}\cdots
u_{i_{k}}^{d_{k}} \rangle.
\end{eqnarray}
{}From Lemma 2 or eq.(3.17), the right side of (3.30) is symmetrized.
Its coefficient is identical to (3.29) by using (3.16), which yields (3.27)
or (3.28).

Thus we have shown that eq.(3.16) is a solution to (3.13). In
turn, it is easy to convert the above verification, particularly that
from eq.(3.23) to (3.25), into an inductive determination of the
expression (3.16) for the coefficient $C$'s starting from the normalized
coefficient at level $l=1$. Thus (3.16) is the unique solution to
eq.(3.13) under the ans\'atz (3.15). If we had set the coefficient of the
leading linear term $u_{m-2}$ to be zero, then all other coefficients in
eq.(3.20) should vanish by induction. (QED)

Now let us go beyond the Ans\'atz (3.15).

Proposition 5: Eq.(3.16) is the only solution to eq.(3.13) with the
most general form (3.8) for charge densities.

Proof: According to eqs.(3.4) and (3.6), any two terms of the same
partition but with different permutations of currents and different
orderings of local products are equal to each other, up to terms
with more derivatives on currents of lower degrees (spins);
in particular, any term can be expressed as the symmetrized
multiple product (3.7) of the same set of currents involved plus terms
with more derivatives. Therefore, we can always rewrite eq.(3.8) as
\begin{eqnarray}
H_{m} &=& \sum_{l}\sum_{\{i\}}C_{i_{1}i_{2}\cdots i_{l}}(m) \langle u_{i_{1}}
u_{i_{2}}\cdots u_{i_{l}}\rangle + \sum_{l}\sum_{\{i,a\}}
C_{i_{1}i_{2}\cdots i_{l}}^{a_{1}a_{2}\cdots a_{l}}(m) \langle
u_{i_{1}}^{(a_{1})}u_{i_{2}}^{(a_{2})}\cdots u_{i_{l}}^{(a_{l})} \rangle
\nonumber\\
&\equiv& H_{m}^{(0)} + P_{m}
\end{eqnarray}
with both $H_{m}^{(0)}$ and $P_{m}$ homogeneous and of degree $m$,
and at least one $a_{k}\neq 0$ in $P_{m}$. We require eq.(3.13):
\begin{eqnarray}
{[\oint_{0} H_{3}^{(0)}(z)dz + \oint_{0} P_{3}(z)dz,
\oint_{0} H_{m}^{(0)}(w)dw + \oint_{0} P_{m}(w)dw ]} = 0.
\end{eqnarray}
Obviously, $H_{3}^{(0)}= u_{1}(z)$, and $P_{3}=0$
up to total derivatives. Also note that the verification of the
coefficients $C_{i_{1}i_{2} \cdots i_{l}}(m)$ in $H_{m}^{(0)}$ is indepedent
of the presence of $P_{m}$ with higher derivatives, so they are indentical
to eq.(3.16). Eq.(3.32) then reduces to
\begin{eqnarray}
{[\oint_{0} H_{3}^{(0)}(z)dz, \oint_{0} P_{m}(w)dw]} =
\oint_{0}dz\oint_{z}dw u_{1}(z) P_{m}(w) = 0.
\end{eqnarray}
We want to show that the only solution to eq.(3.33) is
\begin{eqnarray}
\oint_{0} P_{m}(z)dz = 0.
\end{eqnarray}

Let us prove eq.(3.34) by induction with respect to the number of derivations
on currents: $a=\sum_{j=1}^{l}a_{j}$, and the level $l$.
First we rearrange terms by doing integration by parts so that they
appear with the highest derivative on the highest-spin current being
minimized, e.g., $u_{i_{k}}^{(4)}u_{i_{k}}^{(2)} \rightarrow u_{i_{k}}^{(3)}
u_{i_{k}}^{(3)}$. After doing so, terms with different partitions now
become independent to each other, or irreducible as we call it.
Now let us express terms corresponding to a generic
partition in the form $(u_{i_{1}}^{(a_{1})})^{d_{1}}
(u_{i_{2}}^{(a_{2})})^{d_{2}}\cdots (u_{i_{k}}^{(a_{k})})^{d_{k}}$ with
$\sum^{k}_{j=1}d_{j}i_{j}=l$, $i_{1}\leq i_{2}\leq \cdots \leq i_{k}$ and
$a_{j}<a_{j+1}<\cdots <a_{k}$ for $i_{j}=i_{j+1}=\cdots =i_{k} (j<k)$; and
list them according to the ordering of ($i_{k}$, $a_{k}$, $d_{k}$),
each from large to small.
Consider the first nonvanishing term in the list, which
has largest values of $i_{k}, a_{k}$ and $d_{k}$. Denote
the coefficient of this term as $b$. Its contribution to eq.(3.33) reads
\begin{eqnarray}
& & \oint_{0}dz\oint_{z}dw bu_{1}(z)\langle (u_{i_{1}}^{(a_{1})})^{d_{1}}
(u_{i_{2}}^{(a_{2})})^{d_{2}}\cdots (u_{i_{k}}^{(a_{k})})^{d_{k}} \rangle (w)
\nonumber\\
&=& -2\oint_{0}dz\frac{bd_{k}}{i_{k}+1}\langle (u_{i_{1}}^{(a_{1})})^{d_{1}}
(u_{i_{2}}^{(a_{2})})^{d_{2}}\cdots (u_{i_{k}}^{(a_{k})})^{d_{k}-1}
u_{i_{k}+1}^{(a_{k}+1)} \rangle (z) + \cdots
\nonumber\\
&=& -2\oint_{0}dz\frac{(-1)^{a_{k}+1}bd_{k}}{i_{k}+1}
\langle (u_{i_{1}}^{(a_{1})})^{d_{1}}(u_{i_{2}}^{(a_{2})})^{d_{2}}\cdots
((u_{i_{k}}^{(a_{k})})^{d_{k}-1})^{(a_{k}+1)}u_{i_{k}+1} \rangle (z)
\nonumber\\
& & + {\it ~terms~with~other~partitions}.
\end{eqnarray}
Note the first term on the right side has been arranged irreducible with
no derivative on the highest-spin current $u_{i_{k}+1}$. Assuming eq.(3.34)
is true at $a=a$, and further when $a=a+1$ it is true at $l=l$, then there
will be no other term from the left side of eq. (3.33) matching the
partition of the first term of eq.(3.35) at $l=l+1$. It follows
that $b$ must vanish by using (3.33) and successively all terms in the
list vanish, thus (3.34) is true at $a=a+1$. Note (3.34)
is true at $a=0$, since in this case $P_m(z)\equiv 0$ by definition.
Meanwhile, (3.34) is obviously true at $l=1$ for any $a$,
as all $P_m$'s are total derivatives at this level. This finishes
our proof of (3.34). (QED)

Next we proeed to prove that the commutativity (3.13)
of $Q_m$'s with $Q_3$ will guarantee their mutual commutativity (3.9).
A similar situation happened in the literature for the search
of an infinite set of commuting charges for the quantum KdV equation [15].
Essentially this is a consequence of the Jacobi identities
\begin{eqnarray}
{[Q_{3}, {[Q_{m}, Q_{n}]}]} + {[Q_{m}, {[Q_{n}, Q_{3}]}]} +
{[Q_{n}, {[Q_{3}, Q_{m}]}]} = 0.
\end{eqnarray}

Proposition 6: The above constructed chagres satisfy eq.(3.9).

Proof: We recall that both $Q_{m}$ and $Q_{n}$ are homogeneous and of degree
$m-1$ and $n-1$ respectively. Besides, the OPE's in the $\hat{W}_{\infty}$
algebra are homogeneous, so is the commutator $[Q_{m}, Q_{n}]$ with degree
$m+n-2$. Thus, $[Q_{m}, Q_{n}]$ must be an integral of something which is of
the general form (3.8).

On one hand, substituting eq.(3.13) into (3.36), we have immediately
\begin{eqnarray}
{[Q_{3}, {[Q_{m}, Q_{n}]}]} = 0.
\end{eqnarray}
The above-proved uniqueness of the homogeneous solution (3.15) plus (3.16)
assures us that in view of eq.(3.37), the commutator $[Q_{m}, Q_{n}]$
must be proportional to $Q_{m+n-1}$ up to a constant factor:
\begin{eqnarray}
{[Q_{m}, Q_{n}]} = c~ Q_{m+n-1}.
\end{eqnarray}
We note that on the right side the charge density $H_{m+n-1}$ is led
by the linear term $u_{m+n-3}$ and does not involve any term containing
derivatives of currents.

On the other hand, as a general feature of the $\hat{W}_{\infty}$ algebra,
the commutator between densities $H_{m}$ and $H_{n}$, led by $u_{m-2}$ and
$u_{n-2}$ respectively, does not give rise to the desired leading
$u_{m+n-3}$ term or any term with no derivatives on currents, namely
\begin{eqnarray}
u_{r}(z)\cdot u_{s}(w) &=& {\it terms~with~derivatives~on~currents}
\nonumber\\
& & {\it or~in~powers~other~than~} (z-w)^{-1}.
\end{eqnarray}
This is manifest from the classical $\hat{W}_{\infty}$ algebra [3].
We prove that this feature survives quantization by induction. From eq.(2.24),
(3.39) is true for $r=0,1$. Assuming it is
true for $r=r$, we consider the case with $r$ replaced by $r+1$
and write
\begin{eqnarray}
u_{r+1}(z)\cdot u_{s}(w) &=& \frac{p(w)}{z-w} +
{\it ~terms~with~derivatives~on~currents} \nonumber\\
& & {\it or~in~other~powers~of~}(z-w)^{-1}
\end{eqnarray}
where $p(w)$ is a purely non-derivative polynomial of currents (including
$u_{r+s+2}$) of homogeneous degree $r+s$, appearing only in the
$(z-w)^{-1}$ term. Our basic weapon is the associativity of OPE
\begin{eqnarray}
u_{1}(z)\cdot (u_{r}(x)\cdot u_{s}(w)) =
(u_{1}(z) \cdot u_{r}(x))\cdot u_{s}(w).
\end{eqnarray}
{}From the left side, by the induction assumption,
\begin{eqnarray}
u_{1}(z)(u_{r}(x)u_{s}(w)) &=& u_{1}(z)({\it terms~with~derivatives~on~
currents} \nonumber\\
& & {\it or~in~powers~other~than~} \frac{1}{(x-w)}) \nonumber\\
&=& {\it terms~with~at~least~second~order~derivative} \nonumber\\
& & {\it or~two~derivatives~on~currents} \nonumber\\
& & {\it or~in~powers~other~than~} \frac{1}{(x-w)(z-w)};
\end{eqnarray}
and from the right side, using eq.(2.24),
\begin{eqnarray}
(u_{1}(z)u_{r}(x))u_{s}(w) &=& (\frac{-2}{(r+1)}\frac{u_{r+1}'(x)}{(z-x)}+
{\it ~terms~with~derivatives~on} \nonumber\\
& & {\it lower~spin~currents~or~in~other~powers~of~}\frac{1}{z-x})u_{s}(w)
\nonumber\\
&=& \frac{-2}{(r+1)}\frac{p'(w)}{(z-w)(x-w)} + {\it ~terms~with~at~least}
\nonumber\\
& & {\it second~order~derivative~or~two~derivatives~on~currents} \nonumber\\
& & {\it or~in~other~powers~of~}\frac{1}{x-w} {\it ~or~}\frac{1}{z-w}.
\end{eqnarray}
Comparing eqs.(3.42) and (3.43), we obtain $p(w)=0$. By induction,
eq.(3.39) is true for arbitrary $r$.

Therefore, the constant $c$ in eq.(3.38) must be zero, yielding eq.(3.9).
(QED)

Finally, we emphasize that in the above proofs in this section,
nowhere we have used the two boson realization (2.8) or (2.9) of the
$\hat{W}_{\infty}$ currents $u_{r}$.

\vspace{30 pt}
\section{Quantum KP Hierarchy}
\setcounter{equation}{0}
\vspace{5 pt}

Now with the quantum $\hat{W}_{\infty}(p)$ algebra and an infinite set of
involutive quantum $\hat{W}_{\infty}$ charges (at least at $p=1$)
available, it is straightforward to construct a quantum version of
the KP hierarchy in the Hamiltonian form (1.3).

To this end, we use the quantum $\hat{W}_{\infty}$ (2.19) as the
quantum KP Hamiltonian structure and the densities of the quantum
charges $Q_{m}$ given by eqs.(3.15)-(3.16) as corresponding
Hamiltonian functions $H_{m}$. They naturally generate
an infinite set of compatible flows in various times
$t_{m}$ $(m=1,2,\ldots)$:
\begin{eqnarray}
\frac{\partial u_{r}}{\partial t_{m}} = {[u_{r}, Q_{m+1}]}.
\end{eqnarray}
Since the charges $Q_{m}$ are independent of each other by construction,
so are the flows they generate. Secondly, the mutual commutativity
(3.9) of these quantum charges implies that they are
conserved charges of the flows (4.1):
\begin{eqnarray}
\frac{\partial Q_{n}}{\partial t_{m}} = {[Q_{n}, Q_{m+1}]} = 0.
\end{eqnarray}
Then it is straightforward to check that the flows (4.1) are compatible, i.e.,
\begin{eqnarray}
\frac{\partial^{2} u_{r}}{\partial t_{m}\partial t_{n}} &=&
{[\frac{\partial u_{r}}{\partial t_{m}}, Q_{n+1}]} ~=~
{[ {[u_{r}, Q_{m+1}]}, Q_{n+1}]} \nonumber\\
&=& {[ {[u_{r}, Q_{n+1}]}, Q_{m+1}]} ~=~
\frac{\partial^{2} u_{r}}{\partial t_{n}\partial t_{m}},
\end{eqnarray}
where eq.(4.2) and the Jacobi identity among $u_{r}, Q_{m+1}$ and $Q_{n+1}$
have been applied. Note eqs.(4.1)-(4.3) give rise to the quantum
counterparts of (1.3) and (1.6). They are key features for an integrable
system; especially (4.2) ensures the complete integrability of (4.1). Thus,
we may call the infinite set of operator evolution equations (4.1)
the $p=1$ quantum KP hierarchy, and view it as a desired quantum
version of the classical KP hierarchy (1.3) or (1.1).

We present two arguments for justification of the connection to
the classical KP hierarchy. First by comparaing eq.(3.14) with
(3.1) we note that at least the first three quantum charges $Q_m$
($m=2,3,4$) coincide with the value at $p=1$ of the integral of
the first three quantum Hamiltonians with arbitrary $p$, and
the latter reduce to the classical KP Hamiltonians when
$p\rightarrow 0$. Secondly, for arbitrary $p$, it is very likely, we
conjecture, that there exists an infinite set of commuting
quantum $\hat{W}_{\infty}$ charges, which reduce to the
classical charges of the KP Hamiltonians (1.4) at $p=0$
and give rise to eqs.(3.15)-(3.16) at $p=1$; indeed the charges of
the densities (3.1) represent the first three of such charges.
Assuming the existence of commuting quantum charges $Q_{m}(p)$ for
all $m$, we propose the quantum deformation of eq.(1.3) as
\begin{eqnarray}
\frac{\partial u_{r}(z;p)}{\partial t_{m}} = {[u_{r}(z;p), Q_{m+1}(p)]}.
\end{eqnarray}
Using the explicit charge densities (3.1), we can construct
the first three quantum flows of eq.(4.4) with $m=1,2,3$.
They turn out to be, by applying the OPE's (2.23),
\begin{eqnarray}
& & \frac{\partial u_{0}}{\partial t_{1}} = (1-2p)u_{0}',
{}~~ \frac{\partial u_{1}}{\partial t_{1}} = (1-2p)u_{1}',
{}~~ \frac{\partial u_{0}}{\partial t_{2}} = (1-2p)(2u_{1}'+u_{0}''),
\nonumber\\
& & \frac{\partial u_{1}}{\partial t_{2}} = (2-3p)u_{2}'+(1-p)u_{1}''
+\frac{(1+2p)p}{6}u_{0}''' +p(u_{0}u_{0})', \nonumber\\
& & \frac{\partial u_{0}}{\partial t_{3}} = 3(1-2p)(u_{2}'+u_{1}'')
+\frac{(3-4p-p^{2})}{3}u_{0}''' +3p(u_{0}u_{0})'.
\end{eqnarray}
Manipulating the last three equations, we obtain the first
dynamically nontrivial quantum evolution equation for the
current $u_{0}(z;p)$:
\begin{eqnarray}
6(2-3p)\frac{\partial u_{0}'}{\partial t_{3}}
-9\frac{\partial^{2} u_{0}}{\partial t_{2}^{2}} =
(3-p-16p^{2}+18p^{3})u_{0}'''' +18(1-p)p(u_{0}u_{0})''.
\end{eqnarray}
Rescaling $u_0\rightarrow u_0/p$ and taking $p\rightarrow 0$, this
equation reduces to the classical KP equation, as expected.
This equation is thus justified as a quantum deformation of
the classical KP equation and the hierarchy (4.1) a quantum
version of the KP hierarchy.

Next, let us discuss the field theoretical realization of
the quantum KP hierarchy. Its existence is implied by
the free boson realization of our quantum $\hat{W}_{\infty}(p)$ curents,
whose charges generate the quantum KP flows. According to eqs.(2.8)
or (2.9), the KP variables $u_{r}(z;p)$ can be realized as
functions of two bosonic currents $\bar{j}(z)$ and $j(z)$
and their derivatives, so that the flows (4.4) can be realized
in terms of these currents. Alternatively, instead of
performing such reduction, we prefer to define a basic integrable
hierarchy for $\bar{j}$ and $j$, as a quantum deformation of the
classical $\bar{j}$-$j$ hierarchy [1,16], generated by the quantum charges:
\begin{eqnarray}
\frac{\partial \bar{j}}{\partial t_{m}} = {[\bar{j}, Q_{m+1}(p)]}, ~~~~~~
\frac{\partial j}{\partial t_{m}} = {[j, Q_{m+1}(p)]}
\end{eqnarray}
with $Q_{m}(p)$ exactly the same as in eq.(4.1) or (4.4), but expressed in
terms of $\bar{j}$ and $j$. The Hamiltonian structure of this hierarchy is
directly given by the OPE's (2.7) for $\bar{j}(z), j(z)$. From
eq.(4.7), one obtains all the equations of (4.4) by composition.
In this way, we get a realization (or reduction) the quantum KP
hierarchy using two bosonic currents. Though the two variable hierarchy
(4.7) dynamically is much less rich than the original KP hierarchy
for infinitely many independent variables, the existence of such simple
realization for infinitely many KP flows is still an amazing fact.
Incidentally we remind that the classical limit of the
hierarchies (4.7) and (4.4) is recovered by
rescaling $j\rightarrow j/\sqrt{p}$ and $u\rightarrow u/p$
and ${[~,~]}\rightarrow p{\{~,~\}}$, and then taking $p\rightarrow 0$.

Explicitly, the first few flows in eq.(4.7) read
\begin{eqnarray}
& & \frac{\partial \bar{j}}{\partial t_{1}} = (1-2p)\bar{j}',
{}~~~~ \frac{\partial j}{\partial t_{1}} = (1-2p)j', \nonumber\\
& & \frac{\partial \bar{j}}{\partial t_{2}} = (2-3p)\sqrt{p}(\bar{j}j)'
+\frac{(2-3p)\sqrt{p}}{2}(\bar{j}^{2})' +\frac{p\sqrt{p}}{2}(j^{2})'
+(1-p)\sqrt{1-2p}\bar{j}'', \nonumber\\
& & \frac{\partial j}{\partial t_{2}} = (2-3p)\sqrt{p}(\bar{j}j)'
+\frac{p\sqrt{p}}{2}(\bar{j}^{2})' +\frac{(2-3p)\sqrt{p}}{2}(j^{2})'
-(1-p)\sqrt{1-2p}j''.
\end{eqnarray}
In the derivation we have used
\begin{eqnarray}
Q_{1}(p) &=& \oint_{0} (1-2p)\bar{j}j(z) dz, \nonumber\\
Q_{2}(p) &=& \oint_{0} [\frac{(2-3p)\sqrt{p}}{2}(\bar{j}^{2}j+\bar{j}j^{2})
+\frac{p\sqrt{p}}{6}(\bar{j}^{3}+j^{3}) \nonumber\\
& & -\frac{(1-p)\sqrt{1-2p}}{2}(\bar{j}j'-\bar{j}'j)](z)dz.
\end{eqnarray}

Now, let us discuss physical implications of the above-established
quantum KP hierarchy in the comformal $SL(2,R)_{k}/U(1)$ model. We have
taken the advantage of the quantized  $SL(2,R)_{k}/U(1)$ model to generate
the highly nontrivial quantum deformation of $\hat{W}_{\infty}$ in their two
free boson representation, from which we have further proved the existence
of an infinite set of independent and mutually commuting quantum charges,
at least at level $k=p^{-1}=1$.
These charges generate a huge infinite dimensional quantum symmetry
in the model, given by
\begin{eqnarray}
\delta_{m}\bar{j} = \epsilon_{m}{[\bar{j}, Q_{m+1}]}, ~~~~~~
\delta_{m} j = \epsilon_{m}{[j, Q_{m+1}]}
\end{eqnarray}
or, for the composite currents,
\begin{eqnarray}
\delta_{m} u_{r} = \epsilon_{m}{[u_{r}, Q_{m+1}]},
\end{eqnarray}
with $\epsilon_{m}$ the infinitesimal parameters.
We can easily verify that the quantum KP flows (4.4) are
in fact a set of compatible flows invariant under the symmetry
transformations: Namely, by using eq.(4.2),
\begin{eqnarray}
\delta_{m}(\frac{\partial u_{r}}{\partial t_{n}} - {[u_{r}, Q_{n+1}]})
&=& \epsilon_{m}({[\frac{\partial u_{r}}{\partial t_{n}}, Q_{m+1}]}
- {[{[u_{r}, Q_{m+1}]}, Q_{n+1}]}) \nonumber\\
&=& \epsilon_{m}({[{[u_{r}, Q_{n+1}]}, Q_{m+1}]}
+ {[{[Q_{m+1}, u_{r}]}, Q_{n+1}]}) \nonumber\\
&=& \epsilon_{m}{[u_{r}, {[Q_{n+1}, Q_{m+1}]}]} ~=~ 0.
\end{eqnarray}

In turn, these quantum symmetry flows maintain the quantum $\hat{W}_{\infty}$
algebra invariant. This can be easily seen by considering the fundamental
OPE's (2.7) between the basic bosonic currents $\bar{j}(z)$ and $j(z)$:
Under the flows (4.7), we have infinitesimally,
\begin{eqnarray}
\frac{\partial}{\partial t_{m}}(\bar{j}(z)j(z')-\frac{1}{(z-z')^{2}})
&=& {[\bar{j}(z), Q_{m+1}]}j(z') + \bar{j}(z){[j(z'), Q_{m+1}]} \nonumber\\
&=& {[\bar{j}(z)j(z'), Q_{m+1}]} ~\sim~ {[\frac{1}{(z-z')^{2}}, Q_{m+1}]}
{}~=~ 0.
\end{eqnarray}
So are the OPE's between the composite quantum $\hat{W}_{\infty}$
currents under flows (4.4). In certain sense, the quantum
KP flows generate ``canonical transformations'' in the model.

In conclusion, some discussions are in order. First, we remark
that the quantum KP hierarchy we have established in this paper is
in the Hamiltonian form (1.3), in which the complete integrability of the
hierarchy appears manifest.
In the classical case, the KP hierarchy is usually written in
the equivalent Lax pair form (1.1),
from which all generalized KdV hierarchies can be obtained via natural
reductions. It would be interesting to see if there exists a
Lax-pair-like form for the quantum KP hierarchy,
based on the quantum KP operator (2.17) we proposed in section 2.

Related to this, an interesting problem is to see if the following
expectation is true or not: i.e. our quantum KP hierarchy (4.4)
would contain all known quantum KdV equations first suggested in ref.[15],
and give rise to quantum deformations of generalized
classical KdV hierarchies by reduction. The completely
integrable [17] quantum KdV equations have been shown to
connect to perturbed conformal minimal models [18] and their charges
to the vacuum singular vector of some nonunitary minimal models [19].
For the quantum KP hierarchy, while its commuting charges appear
as an infinite symmetry in the noncompact conformal
$SL(2,R)/U(1)$ model [13], its direct relevance to (perhaps)
perturbed coset conformal field theories remains to be clarified.

Finally, we have obtained the quantum KP hierarchy through
deforming the second classical Hamiltonian structure -- the
nonlinear $\hat{W}_{\infty}$.
It should be possible to obtain a quantum deformation of
the classical KP hierarchy through deforming its much simpler
first Hamiltonian structure -- the $W_{1+\infty}$. Because of its
linearity, quantization of $W_{1+\infty}$ should be straightforward
(either with or without a field realization). In the KP basis, the
complete structure of
the quantum $W_{1+\infty}$ is neatly manifested by the following OPE's:
\begin{eqnarray}
u_{r}(z)u_{s}(w) &=& \sum^{r}_{l=0}\frac{r!}{(r-l)!}
\frac{u_{r+s-l}(z)}{(z-w)^{l+1}} - \sum^{s}_{l=0}(-1)^{l}\frac{s!}{(s-l)!}
\frac{u_{r+s-l}(w)}{(z-w)^{l+1}} \nonumber\\
& & + \frac{(-1)^{s}cr!s!}{(z-w)^{r+s+2}} + O(z-w).
\end{eqnarray}
The remaining issue is to construct an complete set of infinitely many
involutive quantum charges in accordance to eq.(4.14), in order for the
associated quantum KP hierarchy to be integrable.
However, we feel that the recursion relation (see for example [5,3,1])
between the first and second classical KP Hamiltonian structures,
or the bi-Hamiltonian structure, could not survive quantization.

\vspace{30 pt}
\begin{center}
{\bf Acknowledgement}
\end{center}

The work was supported in part by NSF grant PHY-9008452.

{\it Note Added:} After completing the work, we learned that J.
Lukierski and his collaborator also attempted to quantize the KP
equation (private communication).

\vspace{40 pt}
\begin{center}
{\large REFERENCES}
\end{center}
\begin{itemize}
\vspace{5 pt}

\item[1.] F. Yu and Y.-S. Wu, Utah preprint UU-HEP-92/7, Aug. 1992.
\item[2.] M. Sato, RIMS Kokyuroku 439 (1981) 30; E. Date, M. Jimbo, M.
Kashiwara and T. Miwa, in Proc. of RIMS Symposium on Nonlinear Integrable
Systems, eds. M. Jimbo and T. Miwa, (World Scientific, Singapore, 1983);
G. Segal and G. Wilson, Publ. IHES 61 (1985) 1.
\item[3.] F. Yu and Y.-S. Wu, Nucl. Phys. B373 (1992) 713.
\item[4.] L. Dixon, J. Lykken and M. Peskin, Nucl. Phys. B235 (1989) 215.
\item[5.] L. A. Dickey, Annals New York Academy of Sciences, 491 (1987) 131.
\item[6.] F. Yu and Y.-S. Wu, Phys. Rev. Lett. 68 (1992) 2996.
\item[7.] E. Witten, Phys. Rev. D44 (1991) 314.
\item[8.] Y. Watanabe, Ann. di Mat. Pura Appl. 86 (1984) 77.
\item[9.] C. Pope, L. Romans and X. Shen, Phys. Lett. B236 (1990) 173;
B242 (1990) 401.
\item[10.] F. Yu and Y.-S. Wu, Phys. Lett. B263 (1991) 220;
K. Yamagishi, Phys. Lett. B259 (1991) 436.
\item[11.] A. Gerasimov, A. Marshakov and A. Morozov, Nucl. Phys. B328
(1989) 664.
\item[12.] I. Bakas and E. Kiritsis, Maryland/Berkeley/LBL preprint
UCB-PTH-91/44, LBL-31213 or UMD-PP-92/37, Sept. 1991.
\item[13.] F. Yu and Y.-S. Wu, Utah preprint UU-HEP-92/11, May 1992, to be
published in Phys. Lett. B.
\item[14.] O. Hern\'andez, Phys. Lett. 233B (1989) 355.
\item[15.] R. Sasaki and I. Yamanaka, Commun. Math. Phys. 108 (1987) 691;
Adv. Stud. in Pure Math. 16 (1988) 271.
\item[16.] D. A. Depireux, Laval preprint LAVAL-PHY-21-92;
J. M. Figueroa-O'Farrill, J. Mas and E. Ramos,
preprint BONN-HE-92/17, US-FT-92/4 or KUL-TF-92/26.
\item[17.] B. Feigin and E. Frenkel, RIMS/Harvard preprint RIMS-827;
P. Di Francesco, P. Mathieu and D. S\'en\'echal,
Princeton/Laval preprint PUPT-1300 or LAVAL-PHY-28/91.
\item[18.] B. A. Kuperschmidt and P. Mathieu, Phys. Lett. B227 (1989) 245;
T. Eguchi and S. K. Yang, Phys. Lett. B224 (1989) 373;
T. J. Hollpwood and P. Mansfield, Phys. Lett. B226 (1989) 73.
\item[19.] T. Eguchi and S. K. Yang, Phys. Lett. B335 (1990) 282;
A. Kuniba, T. Nakanishi and J. Suzuki, Nucl. Phys. B356 (1991) 750;
P. Di Francesco and P. Mathieu, Phys. Lett. B278 (1992) 79.

\end{itemize}

\end{document}